\documentclass[aps,pre,twocolumn,groupedaddress,showpacs,floatfix]{revtex4}
\usepackage{amsmath}
\usepackage{amssymb}
\usepackage{graphicx}

\begin{document}

\title{Learning about Knowledge: A Complex Network Approach}

\author{Luciano da Fontoura Costa} 
\affiliation{Instituto de F\'{\i}sica de S\~ao Carlos. 
Universidade de S\~ ao Paulo, S\~{a}o Carlos,
SP, PO Box 369, 13560-970, 
phone +55 16 3373 9858,FAX +55 16 3371
3616, Brazil, luciano@if.sc.usp.br}

\date{9th Oct 2005}

\begin{abstract}   
This article describes an approach to modeling knowledge acquisition
in terms of walks along complex networks.  Each subset of knowledge is
represented as a node, and relations between such knowledge are
expressed as edges. Two types of edges are considered, corresponding
to free and conditional transitions.  The latter case implies that a
node can only be reached after visiting previously a set of nodes (the
required conditions).  The process of knowledge acquisition can then
be simulated by considering the number of nodes visited as a single
agent moves along the network, starting from its lowest layer. It is
shown that hierarchical networks, i.e. networks composed of successive
interconnected layers, arise naturally as a consequence of
compositions of the prerequisite relationships between the nodes.  In
order to avoid deadlocks, i.e. unreachable nodes, the subnetwork in
each layer is assumed to be a connected component.  Several
configurations of such hierarchical knowledge networks are simulated
and the performance of the moving agent quantified in terms of the
percentage of visited nodes after each movement.  The
Barab\'asi-Albert and random models are considered for the layer and
interconnecting subnetworks.  Although all subnetworks in each
realization have the same number of nodes, several
interconnectivities, defined by the average node degree of the
interconnection networks, have been considered.  Two visiting
strategies are investigated: random choice among the existing edges
and preferential choice to so far untracked edges.  A series of
interesting results are obtained, including the identification of a
series of plateaux of knowledge stagnation in the case of the
preferential movements strategy in presence of conditional edges.

\end{abstract}

\maketitle

\emph{Knowledge must come through action.}
(Sophocles)
\vspace{0.5cm}

\emph{It is no good to try to stop knowledge from going forward.}
(Enrico Fermi)
\vspace{0.5cm}

\emph{What must the world be like in order that man may know it?} 
(Thomas Kuhn)

\section{Introduction}

Science is the art of building good models of nature, including
science itself.  This is the subject of the present article, i.e. to
revisit the problem of modeling how knowledge is represented and
acquired in the new light of complex network research.

Modeling involves representations of the phenomenon of interest as
well as the dynamics unfolding in such representations in a way which
should be systematically consistent with repetitive confrontation with
experimental data.  Because of their generality, complex
networks~\cite{Albert_Barab:2002,Dorog_Mendes:2002,Newman:2003}
provide a natural and powerful resource for representing structures of
knowledge, where facts are expressed as nodes and relations between
facts are indicated by edges.  Such an approach allows the process of
knowledge acquisition to be modeled in terms of the number of nodes
(or edges) visited during walks through the knowledge network
representation. The present work describes a simple approach to
knowledge acquisition based on complex networks and single-agent
random walks.

The plan of the article is as follows.  After revising the main
related works, focusing on knowledge representation and random walks
in scale free networks, each of the hypotheses adopted in our model
are justified and discussed.  Among other issues, it is shown that
hierarchical networks arise naturally as a consequence of composition
of the prerequisites implied by the conditional links.  Absence of
deadlocks (in the sense of node unreachability) in conditional
transitions are avoided by providing that the networks at each layer
correspond to a connected component(i.e. any node in a layer can be
reached through paths from any other nodes in that same network).
Hierarchical complex networks
(e.g.~\cite{Costa_vor:2003,Costa_topo:2003}) include a series of
layers, each containing a subnetwork, which are interconnected through
subnetworks.  In the proposed model, conditional links are restricted
to those connections between successive layers.  Two types of random
walks are considered, involving random transitions as well as
transitions favoring new links.  The simulations and respectively
obtained results are presented and discussed next, followed by the
development of an analyitical model of plateaux formation.  The
article concludes by presenting several perspectives for further
developments.

\section{A Brief Review of Related Concepts and Developments}

The subject of knowledge representation provides one of the main
issues in artificial intelligence
(e.g.~\cite{Russell:2002,Jackson:1985,Sowa:1999}).  Several discrete
structures, including graphs and trees, have been considered for the
representation of knowledge.  Of particular interest are
\emph{semantic networks}, which code each concept as a node and 
several relationships between such elements (e.g. proximity,
precedence, relative position, etc.) are encoded as edge labels.
However, such structures are mainly considered as a reference for
inferences during pattern analysis, not as a substrate over which to
perform walks or explorations.  The possibility to connect nodes
through logical expressions associated to nodes has provided one of
the main features of random Boolean
networks~\cite{Gershenson:2004,Iguchi:2005}.  These expressions have
been used mainly to combine local states of the nodes, not to control
random walks.  The possibility to associate control on the flow
between nodes in graphs has been adopted in Petri nets
(e.g.~\cite{Reisig:2001}), which has been used mainly for simulating
computing and logical circuits.  The subject of random walks itself
corresponds to a well-developed and important area in statistical
mechanics (e.g.~\cite{Havlin:2000}).  The analysis of random walks in
scale-free networks has been addressed by Tadic
in~\cite{Tadic:2001,Tadic:2003} regarding a special type of network
aimed at simulating the Web, and by Bollt and
Avraham~\cite{Bollt_Avraham:2004} and Noh and
Rieger~\cite{Noh_Rieger:2004} considering recursive and hierarchical
scale-free networks, the latter being concerned with the deterministic
type of hierarchical network proposed in~\cite{Ravasz:2002}.  Random
hierarchical networks similar to those considered in the present work
have been introduced in~\cite{Costa_vor:2003,Costa_topo:2003}.

\section{Hypotheses}

\emph{Representability of Knowledge as a Network:} The basic
assumption underlying the present work is that knowledge can be
represented as a complex network. First, it is understood that
knowledge can be partitioned into chunks which are henceforth
represented as network nodes, while relations between such subsets are
represented as edges.  Two types of edge transitions are considered in
this work: free and conditional.  In the first case, one is allowed to
move freely from a node to the neighboring node, and to come back.
The latter type of transition requires the moving agent to have passed
first along a set of nodes which represent the condition for the
movement.  The process of learning can then be modeled in terms of the
number of nodes (or edges) during walks proceeding along the
respective knowledge network.

\begin{figure}[h]
 \begin{center} 
   \includegraphics[scale=0.5,angle=0]{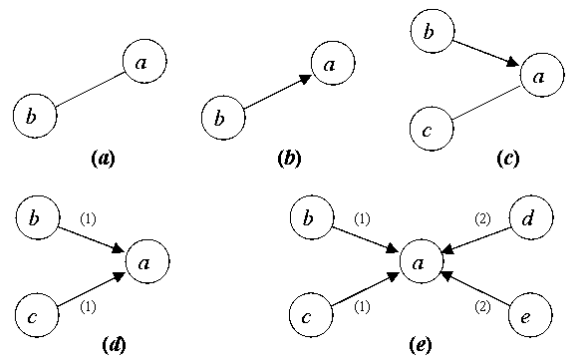} \\
   \vspace{0.5cm} 
   \caption{Types of relationships between knowledge
   subsets (i.e. nodes): equivalence (a); implication (b); hybrid
   relationship involving equivalence and implication (c); multiple
   implication (d) and alternative implications (e).~\label{fig:repr}}
\end{center}
\end{figure}

Figure~\ref{fig:repr}(a) illustrates a free transition between two
subsets $a$ and $b$ of knowledge, while the example in (b) expresses
the simplest conditional case where the moving agent can go from $b$
to $a$, but not from $a$ to $b$, unless it has already made at least
one move from $b$ to $a$.  It is also possible to have hybrid
situations such as those depicted in Figure~\ref{fig:repr}(c), where
$a$ can be reached from $b$ or $c$, but only $c$ can be reached from
$a$.  In order to allow the representation of multiple conditions
(i.e. the fact that $a$ can only be reached after visiting $b_1, b_2,
\ldots, b_A$, $A$ being a positive integer) we introduce the concept
of \emph{token controlled network}.  This multiple conditional case is
illustrated in Figure~\ref{fig:repr}(d).  Here, the subset of
knowledge $a$ can only be reached after visiting $b$ \emph{and} $c$
(in any order).  In other words, it is as if the agent would be
collecting a token from each of the required nodes, which it keeps
henceforth as keys allowing them to proceed through the respective
conditional nodes.  In the present work, it is assumed that all
conditional nodes, i.e. those having tokens required for movement to a
node $n$, are connected to node $n$ through directed edges.  It is
also possible to have alternative multiple conditions, as illustrated
in Figure~\ref{fig:repr}(e), where the labels associated to the edges
identify the respective conditional structures.  In this case, $a$ can
be reached if and only if both $b$ and $c$ were visited before; or
after visiting $d$ and $e$.  The case in which a node $a$ can be
accessed after visiting $b$ or $c$ is represented by two undirected
edges, without associated labels, from those nodes to $a$.  In brief,
the free edges are represented by undirected edges, and the
conditional by directed arrows with associated identifying labels.
Alternative multiple conditions are not considered in the present work
in order to limit the complexity and number of parameters in the
experimental and analytical characterization of the dynamics of
knowledge acquisition.

Regarding the movement of agents along such networks as they integrate
the knowledge available from the nodes, it is natural that a free
transition can be tracked in any direction.  However, a conditional
edge from $b$ and $c$ to $a$ is considered to be direction restrictive
only until $a$ is reached for the first time (after visiting $b$ and
$c$), becoming a free edge henceforth.  This type of dynamics is
implemented in order to express the fact that once knowledge about
$a$, $b$ and $c$ is achieved, i.e. the conditional transition is
mastered, it becomes possible to reach any of the conditions from node
$a$.  

\begin{figure}[h]
 \begin{center} \includegraphics[scale=0.62,angle=0]{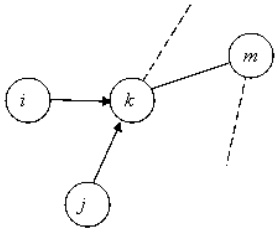} \\
   \vspace{0.5cm} \caption{The indiscriminate use of multiple
   conditional implications quickly leads to deadlocks such as that
   illustrated here.  The subset of knowledge in node $k$ can never be
   reached by an agent starting at $i$ or $j$, as there is no
   connection between these two nodes.~\label{fig:deadlock}}
\end{center}
\end{figure}

\emph{Hierarchical Knowledge Networks:} The indiscriminate
incorporation of the multiple conditions into a network can easily
lead to deadlocks such as that illustrated in
Figure~\ref{fig:deadlock}.  This deadlock is a direct consequence of
the fact that there is no path connecting $i$ and $j$ in the network
represented in this figure.  We henceforth assume that the knowledge
network is \emph{consistent}, in the sense that all nodes should be
reachable.  In addition, the several prerequisites between the
portions of knowledge assigned to nodes naturally define a hierarchy
along the networks.  For instance, some knowledge at node $i$ may
require previous visits to nodes $j$ and $k$, which we shall represent
as $i=P(j,k)$.  The access to $k$ and $j$ may demand previous visits
to nodes $m$ and $n$ and $q$ and $r$, respectively, represented by the
composition of prerequisites $i=P(j,k)=P(P(m,n),P(q,r))$, implying two
hierarchies.  Therefore, a network of consistent knowledge can be
general and naturally organized as a \emph{hierarchy} of $H$ layers,
with the first layer corresponding to all nodes which have no
prerequisites, while the remainder of the nodes are partitioned into
layer by the composed prerequisites. Each layer $h$ contains a
connected subnetwork (i.e. any node in the subnetwork can be reached
through at least one path from any node) which is interconnected, via
conditional edges, to nodes in layer $h-1$.  It is important to note
that, although related to previous works such
as~\cite{Costa_vor:2003,Costa_topo:2003}, the hierarchical
organization for knowledge representation is self-contained and
follows naturally from knowledge consistence and the composition of
prerequisites between nodes.  The total number of layers is henceforth
expressed as $H$, the number of nodes in layer $h$ as $N(h)$, while
the number of nodes in the whole hierarchical network is denoted as
$\Omega$.

\begin{figure}
 \begin{center} \includegraphics[scale=0.62,angle=0]{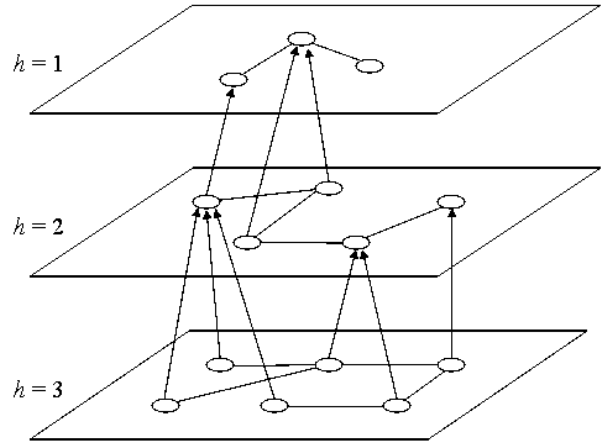} \\
   \vspace{0.5cm} \caption{A example of simple hierarchical 
   network.~\label{fig:hier}}
\end{center}
\end{figure}

Figure~\ref{fig:hier} illustrates a simple hierarchical knowledge
network containing three layers.  For simplicity's sake, hybrid
relationships or alternative implications are not considered
henceforth.  In addition, all \emph{network layers} are assumed to be
of the same type (e.g. random or Barab\'asi-Albert --- BA) and have
the same number of nodes and average node degree.  The nodes at the
highest hierarchy are called assumptions, and are the place where all
the walks start.  Note that the \emph{highest} hierarchical levels are
found at the \emph{lowest} portion of Figure~\ref{fig:hier}.

\begin{figure}
 (a) \includegraphics[scale=0.62,angle=0]{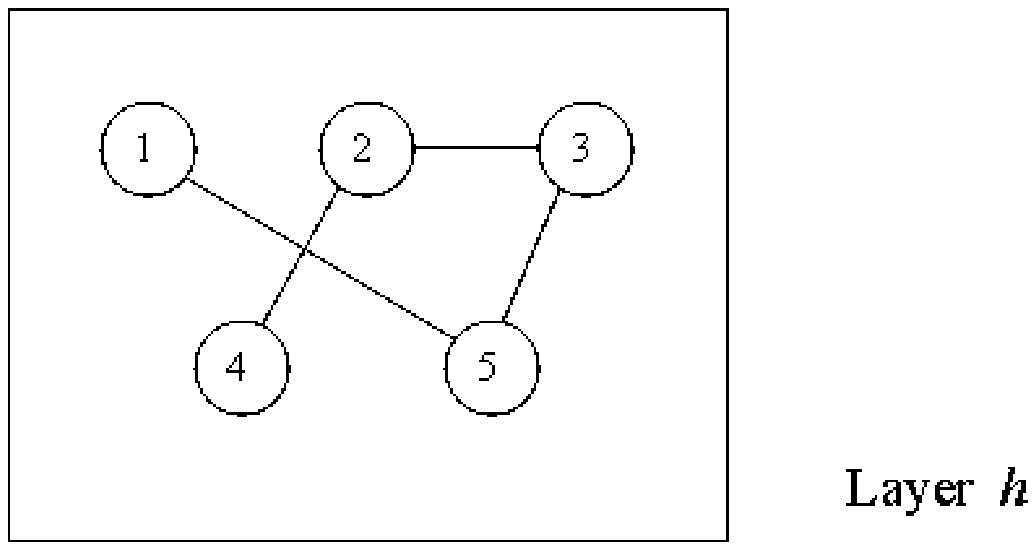} \\
 (b) \includegraphics[scale=0.62,angle=0]{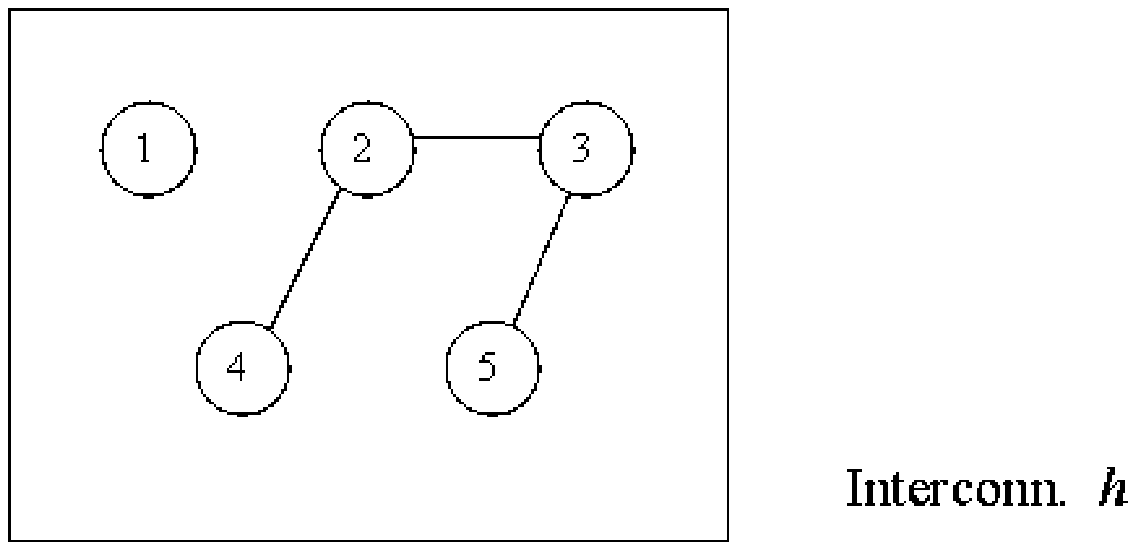} \\
 (c) \includegraphics[scale=0.62,angle=0]{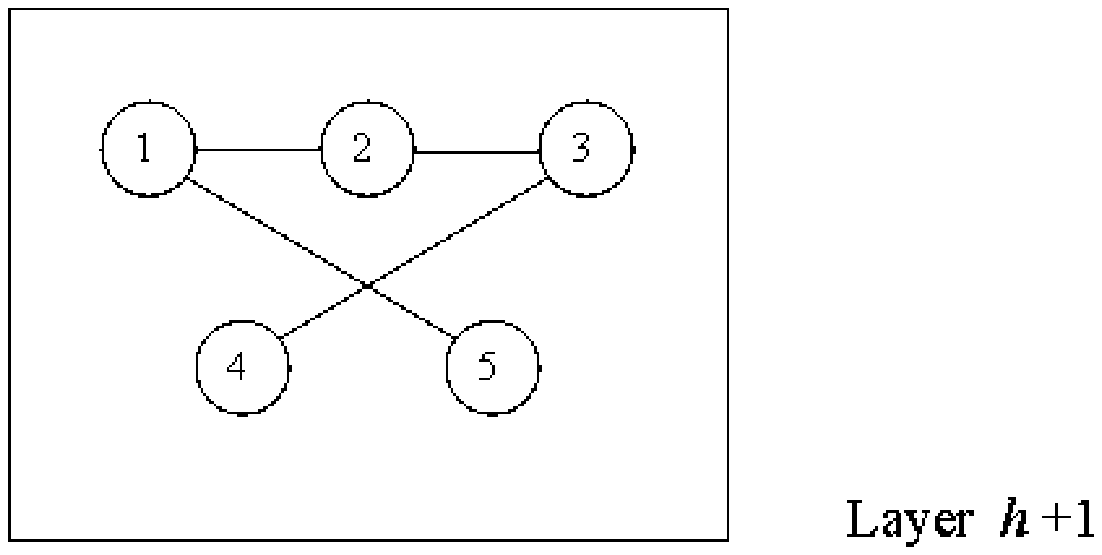} \\
 \begin{center} 
   \includegraphics[scale=0.62,angle=0]{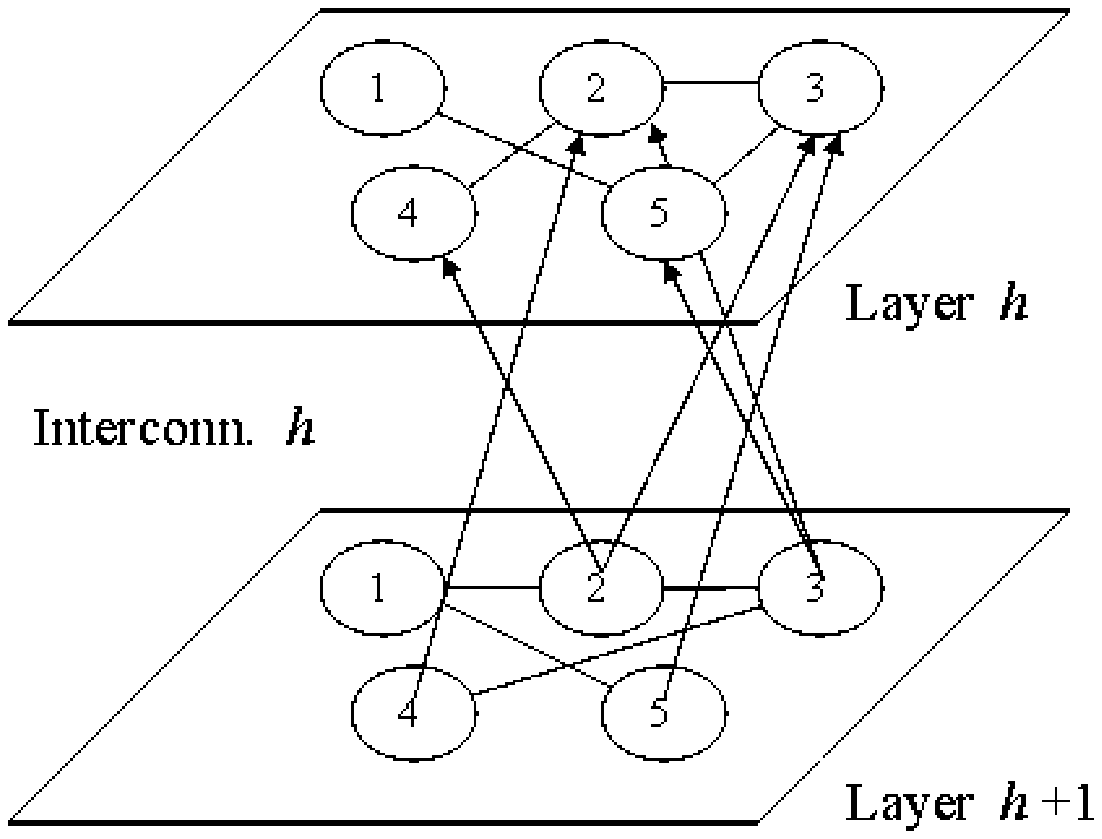} \\
   (d)
   \vspace{0.5cm} 
   \caption{Two layers (a,c) and one interconnecting (b) simple
   subnetworks and the section of the hierarchical network
   respectively implemented (d).~\label{fig:interconn}}
\end{center}
\end{figure}

The set of \emph{interconnecting networks} is also of uniform type and
have the same number of nodes and edges.  In the current work, these
subnetworks can be of random (i.e. Erd\H{o}s-R\'enyi) or BA types,
defining how the subnetwork in layer $h+1$ connects to the nodes in
layer $h$.  Figure~\ref{fig:interconn} illustrates how such
interconnections are henceforth understood.  The layers $h$
(Figure~\ref{fig:interconn}(a)), and $h+1$
(Figure~\ref{fig:interconn}(c))are to be connected through the
interconnection subnetwork $h$ in Figure~\ref{fig:interconn}(b).  Each
edge $(i,j)$ in the interconnection layer implies that node $i$ in
layer $h$ is connected to node $j$ in layer $h-1$ \emph{and} that node
$j$ in layer $h$ is connected to node $i$ in layer $h-1$. Note that
although a more flexible interconnecting scheme could be achieved by
using directed interconnecting networks, the present study considers
all layer and interconnecting networks to be undirected because such a
structure favors the analytical model developed in
Section~\ref{sec:analytic} without loss of generality except for the
respectively implied doubled average node degree.  The connections
implemented by the three subnetworks in
Figure~\ref{fig:interconn}(a-c) are illustrated in
Figure~\ref{fig:interconn}(d).

\section{Computational Implementation}

\begin{figure}
 \begin{center} 
 \includegraphics[scale=0.5,angle=0]{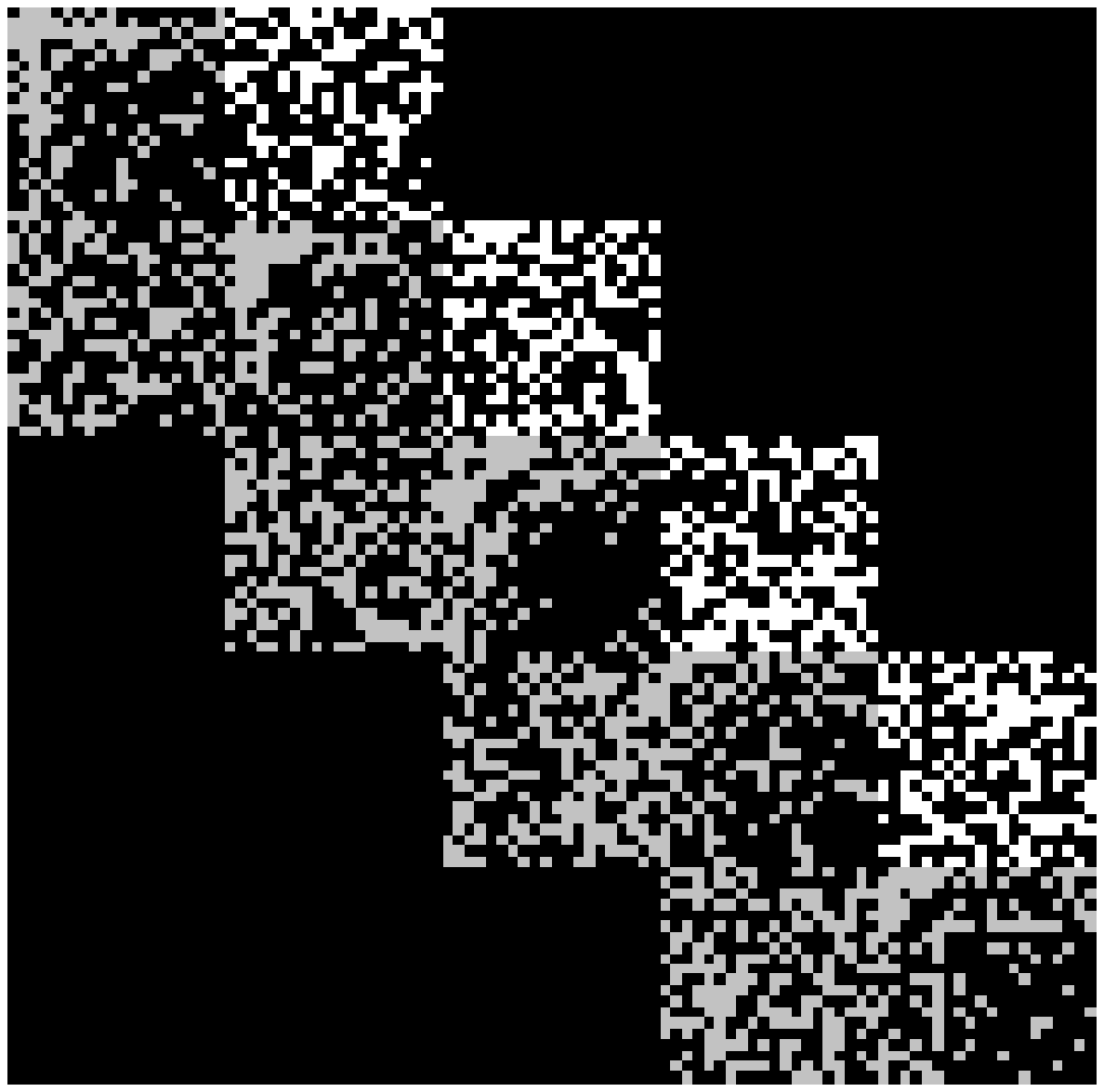} \\
 \vspace{0.5cm} 
 \caption{Example of extended adjacency matrix $K$ considering BA
 layers and random interconnections. The conditional connections are
 represented in white and the equivalence edges in
 gray.~\label{fig:matr}}
\end{center}
\end{figure}

Knowledge networks involving the free and conditional edges described
above can be conveniently represented in terms of an \emph{extended
adjacency matrix}~\footnote{The term \emph{weight matrix} has been
deliberately avoided here because the values (labels) in the matrix
are more related to the adjacency between nodes than to weights.}
henceforth represented as $K$.  Each node is labeled by consecutive
integer values $1, 2, \ldots, N$.  The equivalence between two nodes
$i$ and $j$ is indicated by making $K(i,j)=1$ and $K(j,i)=1$.  The
single conditional connection from node $i$ to $j$ is represented as
$K(i,j)=1$ and $K(j,i)=-1$. Note that such an assignment implements
the adopted strategy that an implication edge can be backtraced
unconditionally.  The multiple conditional transition from $i_1,
i_2,
\ldots, i_A$ to $j$ is represented as $K(i_p,j)=1$ and $K(j,i_p)=-1, p
= 1, 2, \ldots, A$.  Figure~\ref{fig:matr} illustrates an extended
adjacency matrix $K$ considering BA models for layer and
interconnecting networks. 

The moving agent keeps at all times a vector $\vec{v}$ of visited
nodes and an individual adjacency matrix $G$, which are continuously
updated after each movement.  The agent is granted to know about all
feasible connections emanating from the current node $i$, while the
feasibility of a given edge $(i,j)$ is decided by taking into account
its list $\vec{v}$ of visited nodes.  More specifically, an edge will
be feasible, and accessible to the agent, in case it has already
visited the required nodes and collected the respective tokens. 

The movement strategies described in the two following subsections
have been considered in the reported simulations.

\subsection{Random choice of edges} 

In this case, the next edge to be taken from the current node $i$ is
drawn with the same probability between all the feasible connections
between $i$ and all other nodes.  By feasible connection it is meant
either a free edge or a conditional edge for which all conditions have
already been met.

\subsection{Preferential choice of edges} 

Unlike the previous case, the free edges which have not yet been
tracked are considered first, with uniform probability.  In case no
such edges exist, the next edge is drawn uniformly among the other
allowed movements, i.e.  free links which have already been tracked
and enabled conditional links which remain untracked.  The exclusion
of the untracked conditional links, even if enabled, from the
preferential movements is considered in order to express the fact that
such a kind of knowledge enlargement is more demanding than exploring
first the untracked unconditional connections.

Note that in neither case the agent uses its knowledge about the
status (i.e. already visited or not) of the node to which the
emanating edges lead to.  Although more sophisticated moving
strategies which make full use of the information stored in partial
graph $G$ stored with the agent can be devised, including the choice
of shortest paths to unvisited edges, they are not pursued further in
the present work.

The two strategies above aim to represent, though very na\"{\i}vely
and incompletely, two possible ways to acquire knowledge.  In the
first case, no distinction is made between a new or already taken
relation.  It is as if the researcher (i.e. the agent walking through
the network) is equally interested in revising a relationship or
seeking for new possible connections.  In the second visiting scheme,
the researcher is more actively interested in exploring new
relationships, resorting to already tracked connections or enabled
conditional links only in case no untracked free links are available.
Intuitively, the second strategy would seem to be more effective in
finding new knowledge, by covering the edges more effectively.

\section{Simulations}

For simplicity's sake, all simulations reported in this work are
restricted to hierarchical networks with $H=5$ network layers, with
all layer and interconnecting subnetworks having $N(h)=N=20$ nodes,
implying a total of $\Omega=100$ nodes for all layer subnetworks
(larger networks involve much longer execution times).  Random and
Barab\'asi-Albert models are considered for layers and
interconnections.  The latter are defined by the number of edges $m$
of each new added node, starting with $m0$ nodes.  For each such BA
network, an `equivalent' random network -- in the sense of having the
same average degree $\left< k \right>$ and number of edges $N_E$ -- is
obtained.  Since the average degree of a BA network with $m$ edges per
node is known to be

\begin{equation}
  \left< k \right> = 2 m,
\end{equation}

the Poisson rate of the equivalent random (Erd\H{o}s-R\'enyi) network
with the same number of $N$ nodes and same average degree $\left< k
\right>$ can be verified to be given as

\begin{equation}
  \gamma_e = 2m/(N-1).  \label{eq:lamb}
\end{equation}

The above result follows from the fact that in an Erd\H{o}s-R\'enyi
network we have $\gamma=\left< k \right> /(N-1)$.  The number of edges
in any of the BA or random networks can be calculated as

\begin{equation}
  N_E=N/2 \left< k \right>.
\end{equation}

Therefore, in this work the values of $m$ are used to define the
connectivity of the BA models and then of the respective random
counterparts.

Three configurations have been chosen for the BA layer models: $m=1,
5$ and $10$, while eight configurations are considered for the
interconnecting networks: $m = 1, 2, \ldots, 8$.  Table~\ref{tab:meas}
shows the expected values of number of edges $N_E$, average degree
$\left< k
\right>$, and Poisson rate $\gamma_e$ for values of $m$ ranging from 1 to 10.

\begin{table}
  \vspace{1cm} 
\begin{tabular}{||l|l|l|l||} \hline 
$m$ & $N_E$ &  $\left< k \right>$ & $\gamma_e$ \\ \hline 
1 & 20 & 2 & 0.105 \\ 
2 &  40 & 4 & 0.210 \\ 
3 & 60 & 6 & 0.316 \\ 
4 & 80 & 8 & 0.421 \\ 
5 &  100 & 10 & 0.526 \\ 
6 & 120 & 12 & 0.631 \\ 
7 & 140 & 14 & 0.737 \\
8 & 160 & 16 & 0.842 \\ 
9 & 180 & 18 & 0.947 \\ 
10 & 200 & 20 &  1.05 \\ \hline 
\end{tabular} 
\caption{Values of $m$ for the BA
  models and the respective total number of edges $N_E$, average
  degree $\left< k \right>$ and equivalent Poisson rate $\gamma_e$
  expected for each subnetwork (layer or interconnecting) with $N=20$
  nodes.}~\label{tab:meas}
\end{table}

The following configurations were addressed in the reported
simulations:

{\bf (i)} all layers and interconnecting subnetworks are BA; \linebreak
{\bf (ii)} all layers subnetworks are random and all interconnecting
networks are BA; \linebreak
{\bf (iii)} all layers subnetworks are BA and all
interconnecting subnetworks are random; \newline
{\bf (iv)} all layers and interconnecting subnetworks are random.  \newline

Each of the above configurations was investigated while considering
two visiting strategies: (a) allowed edges are chosen randomly; and
(b) if available, untracked allowed edges are selected randomly,
otherwise allowed tracked edges are selected randomly.  In order to
assess the effect of the conditional edges between successive layers,
counterparts of each considered configuration interconnected by
unconditional networks have also been simulated and had their
performance quantified.  Although several alternative or complementary
performance indices could have been considered, for simplicity's sake
our attention is restricted to the percentage $P$ of visited nodes and
percentage $E$ of visited edges at time instant $t$.  The speed of
knowledge acquisition can be estimated by taking the time derivative
of this quantity, i.e. $\dot{P}$.  A total of 100 realizations
involving $N_t=2400$ time steps (corresponding to each movement along
the walk) have been performed.

\begin{figure*}[h]
 \begin{center} 
   \includegraphics[scale=0.62,angle=0]{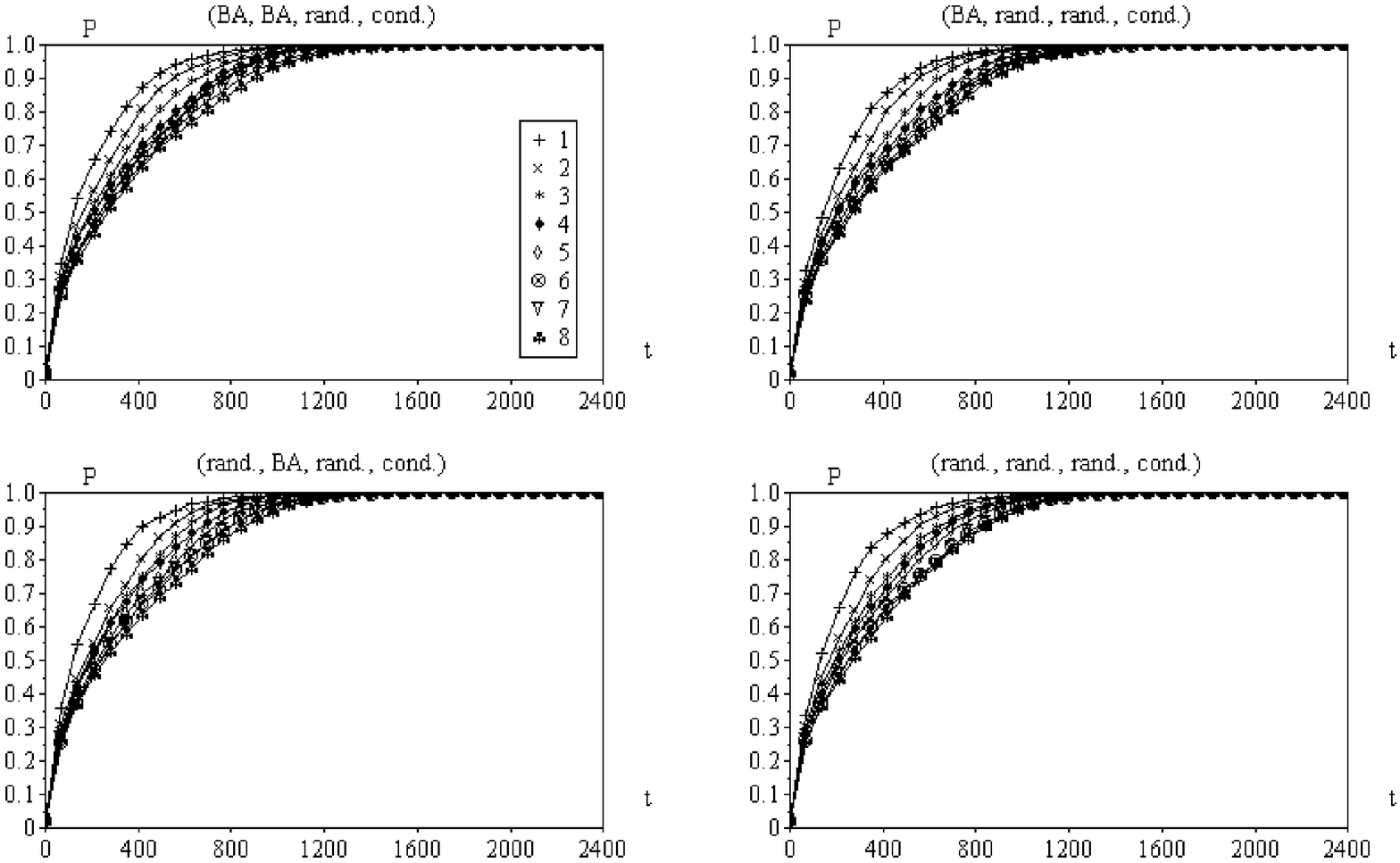} \\
   \vspace{0.5cm} 
   \caption{The percentage of visited nodes $P$ in terms of the time
   $t$ for the configuration involving random choice of edges
   and conditional interconnections. The legend indicates the density
   of interconnections in terms of $m$.~\label{fig:rand_cond}}
\end{center}
\end{figure*}

Figure~\ref{fig:rand_cond} shows the learning curves obtained for
$P(t)$ considering layer networks defined by $m=5$ and five
interconnecting layers with $m=1, 2, \ldots, 8$ for the several
combinations of $layer/interconnection$ types of networks, presence of
conditional connections between layers, and consideration of the
random choice of movement.  The title of each graph is henceforth
organized as $(layer, interconn., movement, conditionality)$, where
$layer$ and $interconn.$ indicates the model assumed for the layer and
interconnecting networks, $movement$ identifies the moving agent
strategy (random or preferential for new edges), and $conditionality$
indicates the type of interconnecting edges (conditional or
free/unconditional).

\begin{figure*}[h]
 \begin{center} 
   \includegraphics[scale=0.62,angle=0]{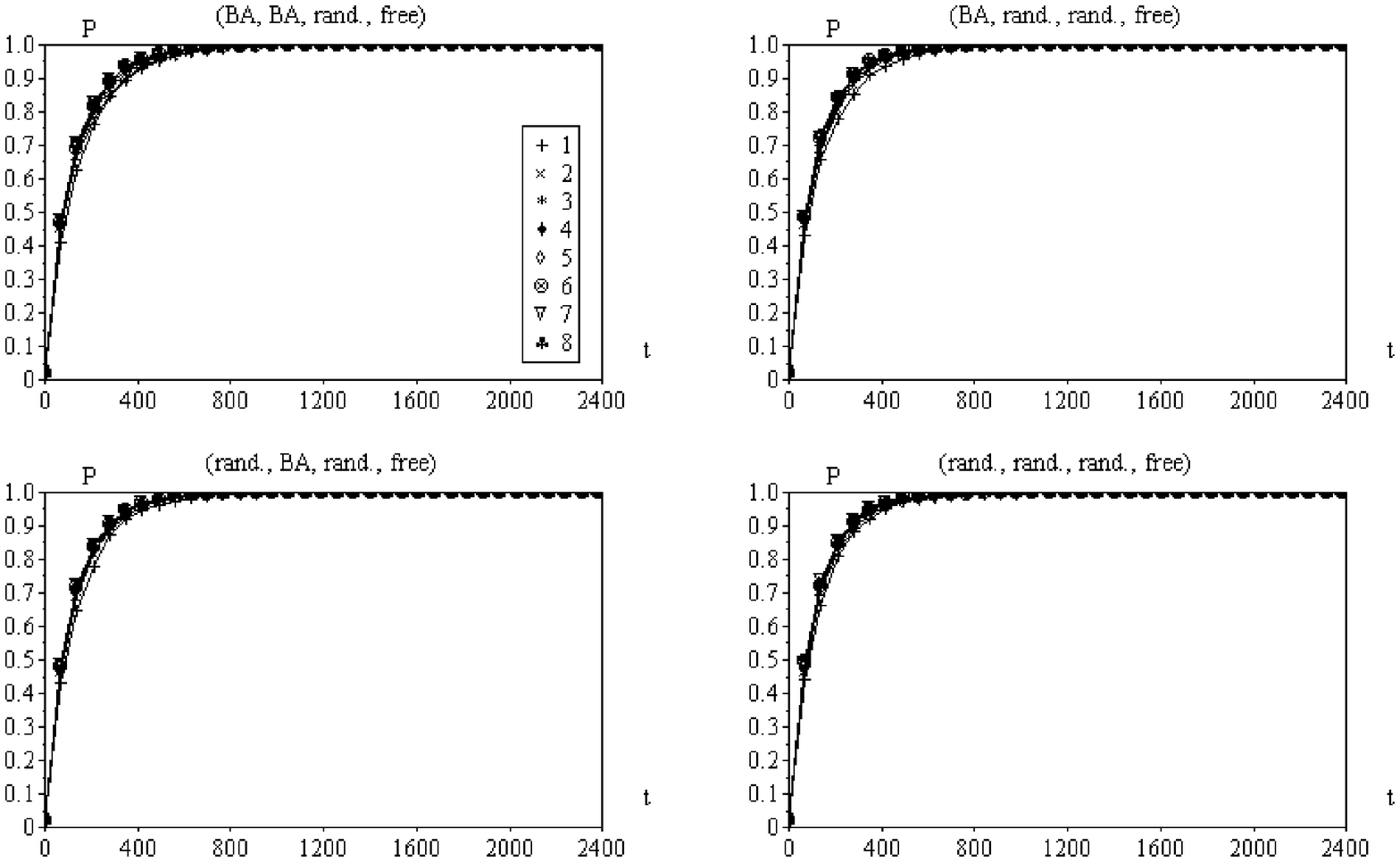} \\
   \vspace{0.5cm} 
   \caption{The percentage of visited nodes $P$ in terms of the time
   $t$ for the configuration involving random choice of edges
   and unconditional interconnections.~\label{fig:rand_uncond}}
\end{center}
\end{figure*}

\begin{figure*}[h]
 \begin{center} 
   \includegraphics[scale=0.62,angle=0]{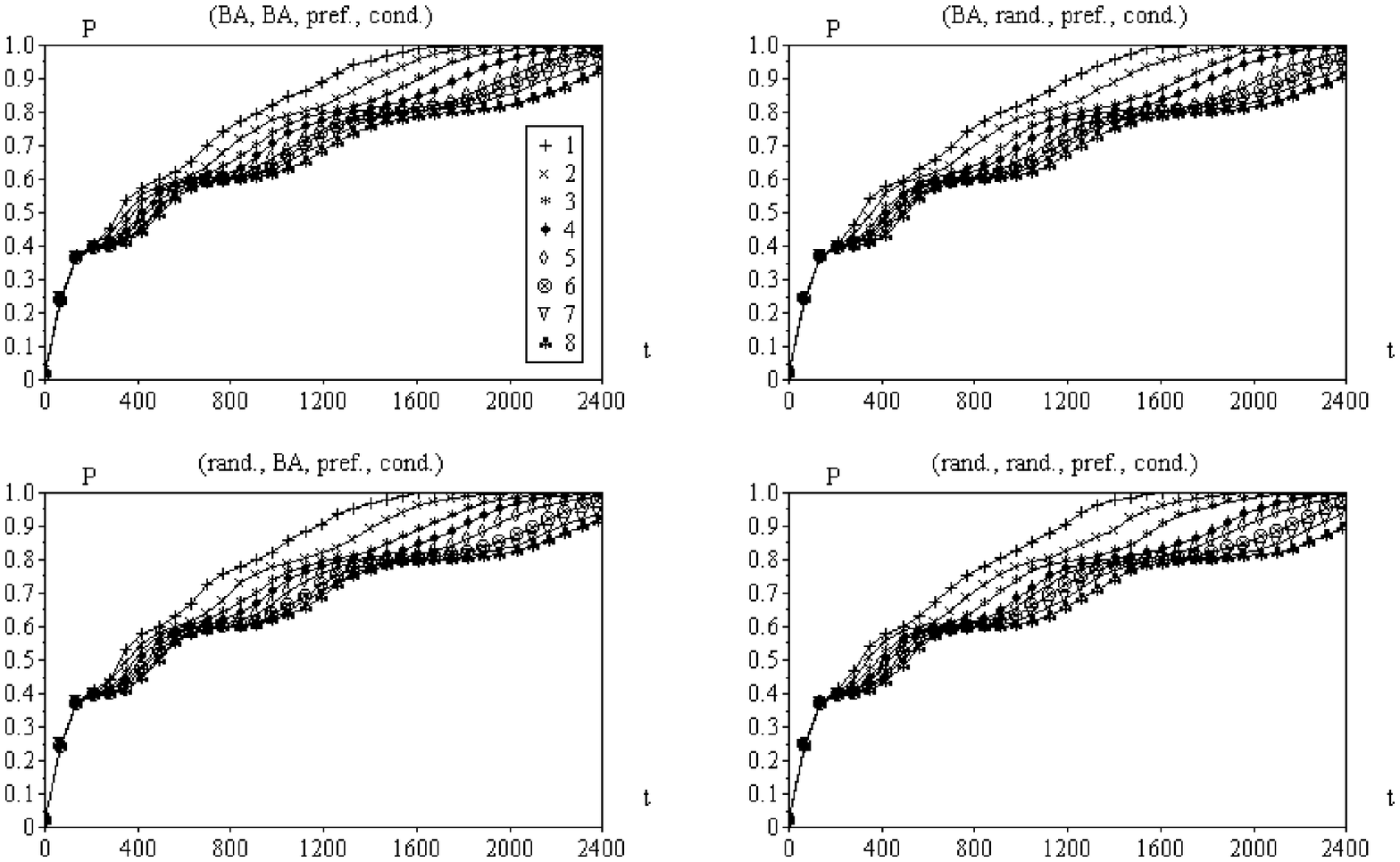} \\
   \vspace{0.5cm} 
   \caption{The percentage of visited nodes $P$ in terms of the time
   $t$ for the configuration involving preferential choice of edges
   and conditional interconnections.~\label{fig:pref_cond}}
\end{center}
\end{figure*}

\begin{figure*}[h]
 \begin{center} 
   \includegraphics[scale=0.62,angle=0]{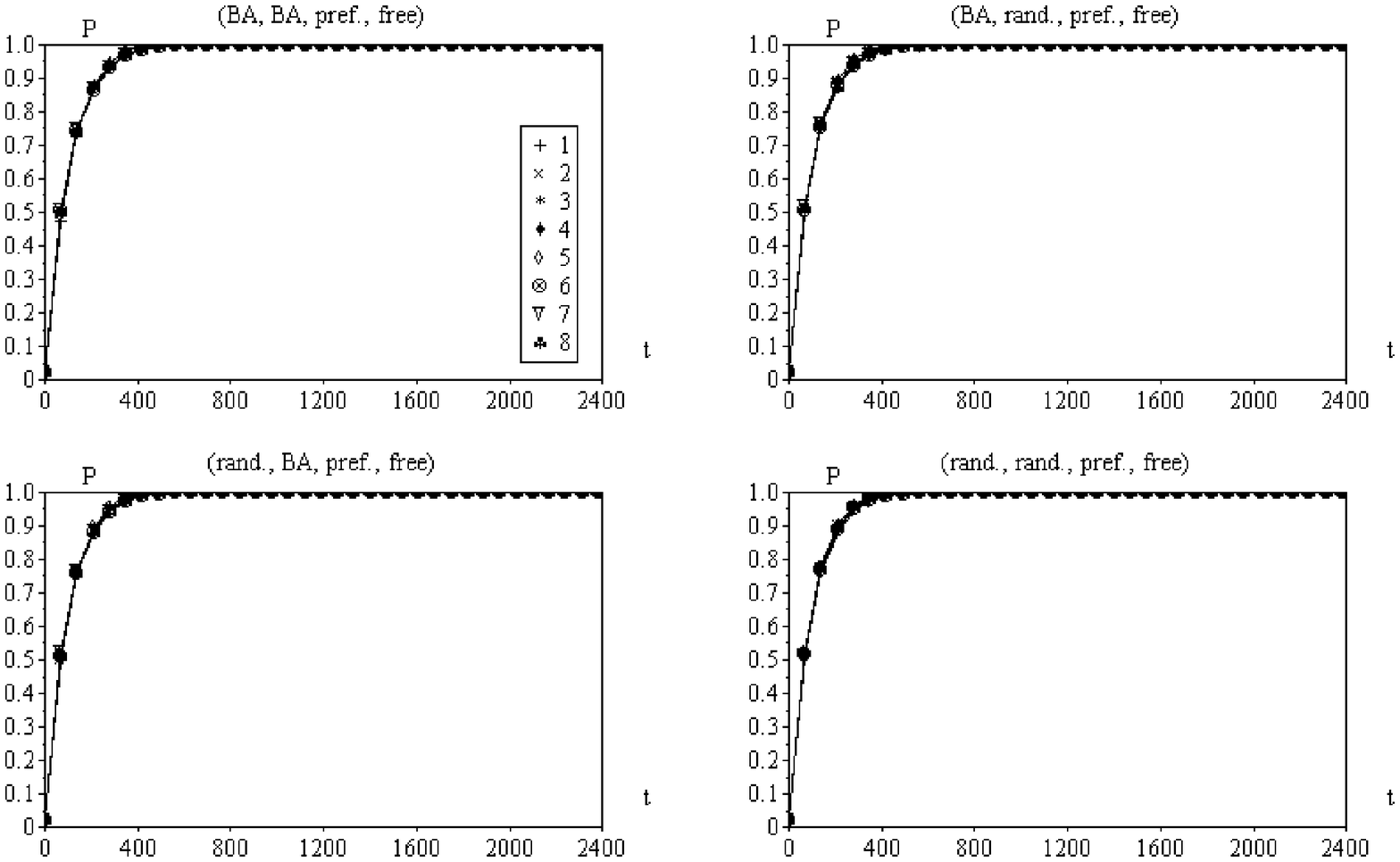} \\
   \vspace{0.5cm} 
   \caption{The percentage of visited nodes $P$ in terms of the time
   $t$ for the configuration involving preferential choice of edges
   and unconditional interconnections.~\label{fig:pref_uncond}}
\end{center}
\end{figure*}

Analogous results obtained for the preferential movements /
conditional connections; random movements / unconditional connections
and preferential movements / unconditional connections are given in
Figures~\ref{fig:rand_cond}, ~\ref{fig:rand_uncond},
~\ref{fig:pref_cond} and ~\ref{fig:pref_uncond} respectively.  In all
the remaining figures in this article, the legend bar indicates the
density of the interconnections.  The values in these legends
correspond to the parameter $m$ adopted for the BA model, therefore
defining the density of interconnections for this model and also for
the equivalent random counterpart (see Equation~\ref{eq:lamb}).

\begin{figure*}[h]
 \begin{center} 
   \includegraphics[scale=0.62,angle=0]{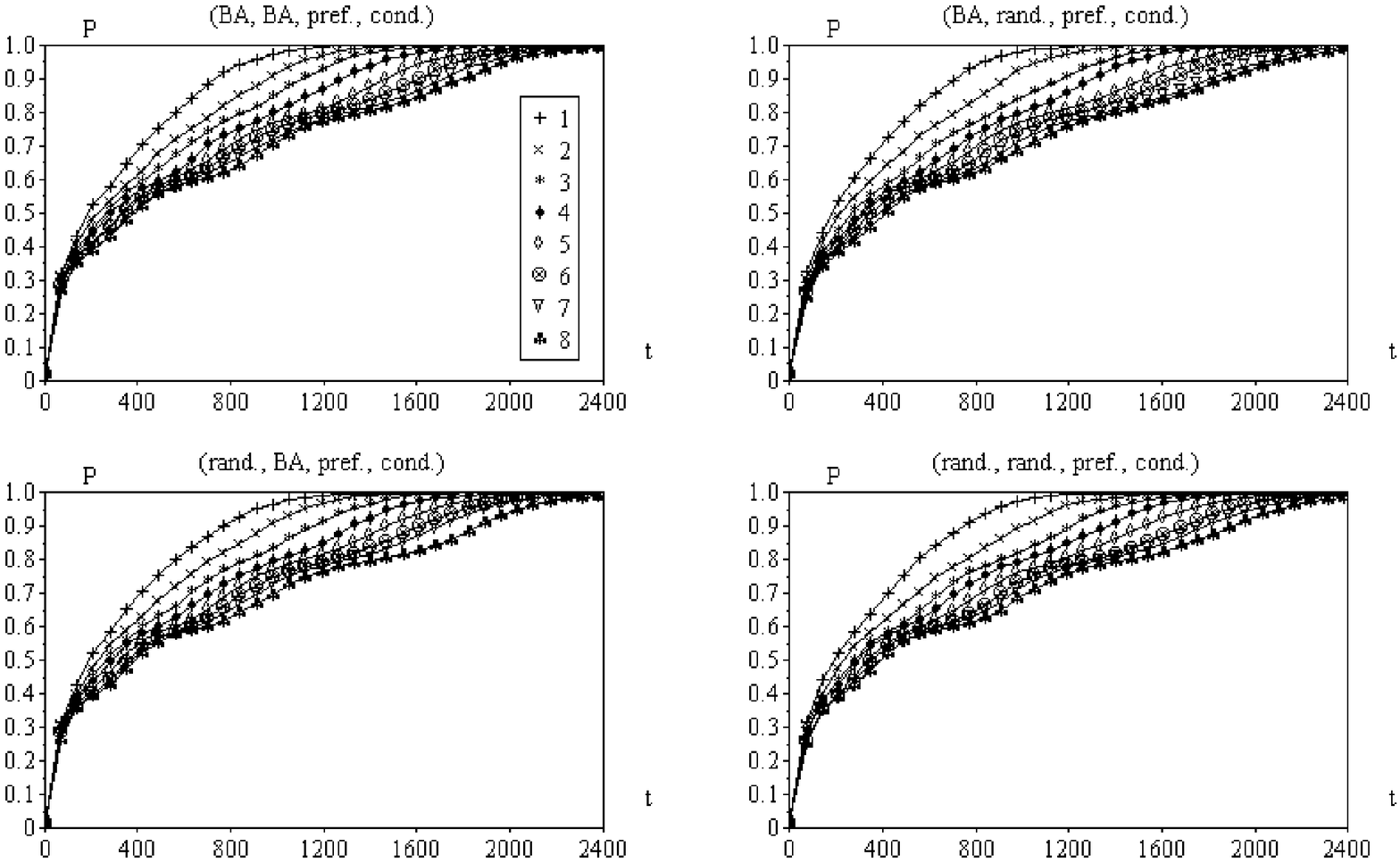} \\
   \vspace{0.5cm} 
   \caption{The percentage of visited nodes $P$ in terms of the time
   $t$ for the configuration involving preferential choice of edges
   and conditional interconnections. All layer networks
   consider $m = 1$~\label{fig:pref_cond_1}}
\end{center}
\end{figure*}

\begin{figure*}[h]
 \begin{center} 
   \includegraphics[scale=0.62,angle=0]{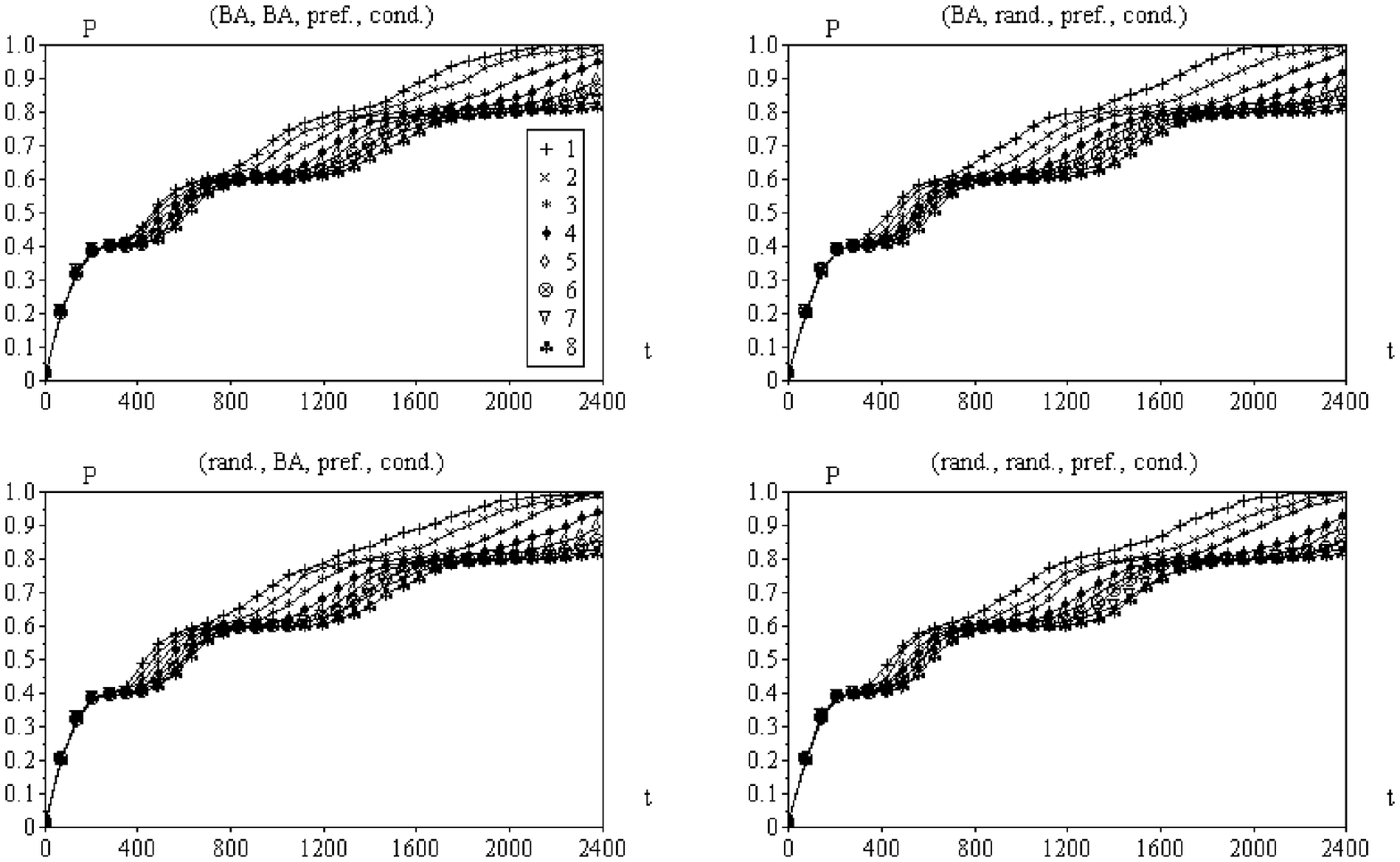} \\
   \vspace{0.5cm} 
   \caption{The percentage of visited nodes $P$ in terms of the time
   $t$ for the configuration involving preferential choice of edges
   and conditional interconnections. All layer networks
   consider $m=10$~\label{fig:pref_cond_10}}
\end{center}
\end{figure*}

Models considering different connectivities for the layer networks,
namely $m=1$ and $m=10$, have also been simulated and investigated.
The percentage of visited nodes $P(t)$ obtained for the preferential
choice / conditional interconnections situation is shown in
Figures~\ref{fig:pref_cond_1} and~\ref{fig:pref_cond_10},
respectively.

\begin{figure*}[h]
 \begin{center} 
   \includegraphics[scale=0.62,angle=0]{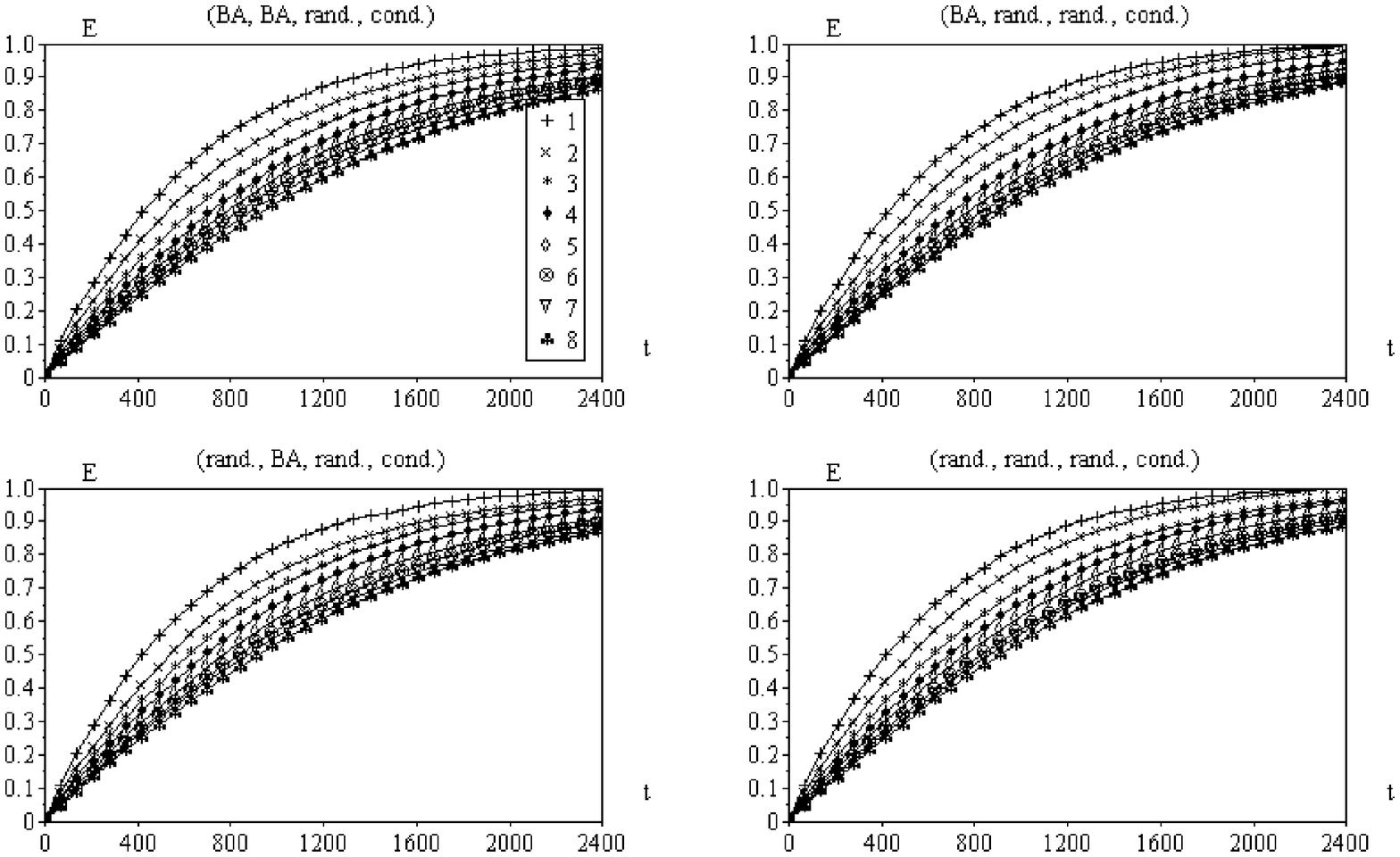} \\
   \vspace{0.5cm} 
   \caption{The percentage of visited edges $E$ in terms of the time
   $t$ for the configuration involving random choice of edges
   and conditional interconnections.~\label{fig:E_rand_cond}}
\end{center}
\end{figure*}

\begin{figure*}[h]
 \begin{center} 
   \includegraphics[scale=0.62,angle=0]{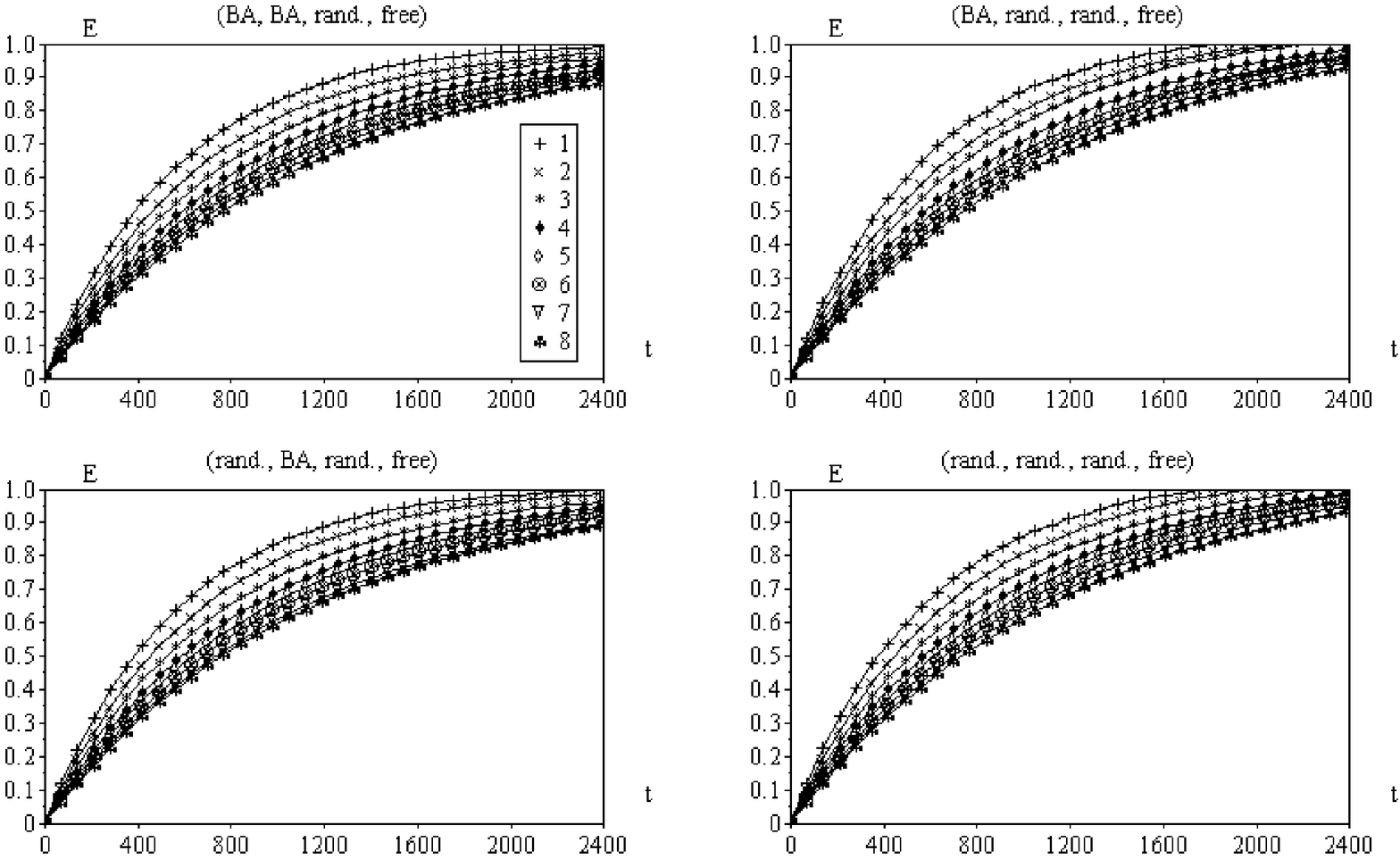} \\
   \vspace{0.5cm} 
   \caption{The percentage of visited edges $E$ in terms of the time
   $t$ for the configuration involving random choice of edges
   and unconditional interconnections.~\label{fig:E_rand_uncond}}
\end{center}
\end{figure*}

\begin{figure*}[h]
 \begin{center} 
   \includegraphics[scale=0.62,angle=0]{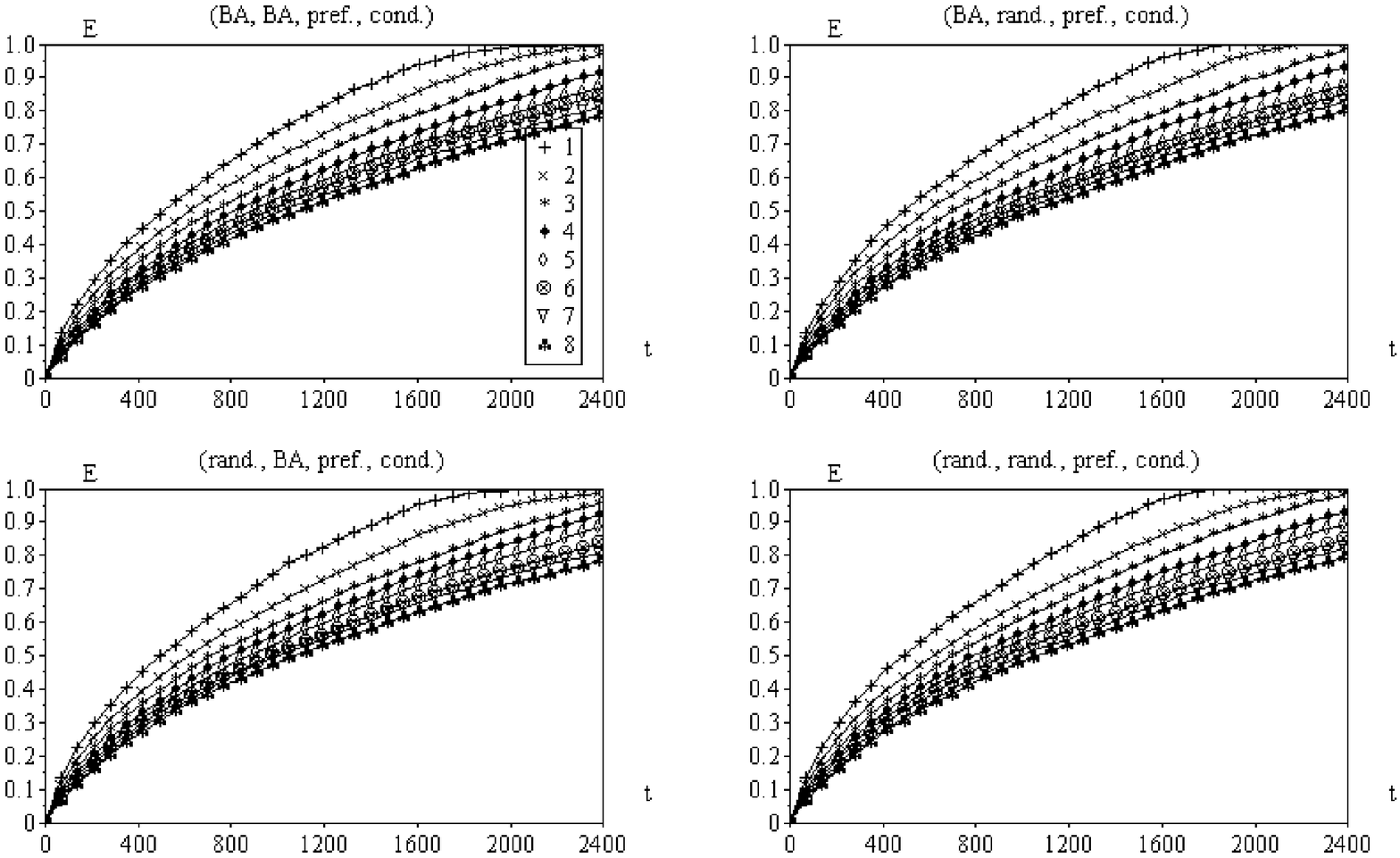} \\
   \vspace{0.5cm} 
   \caption{The percentage of visited edges $E$ in terms of the time
   $t$ for the configuration involving preferential choice of edges
   and conditional interconnections.~\label{fig:E_pref_cond}}
\end{center}
\end{figure*}

\begin{figure*}[h]
 \begin{center} 
   \includegraphics[scale=0.62,angle=0]{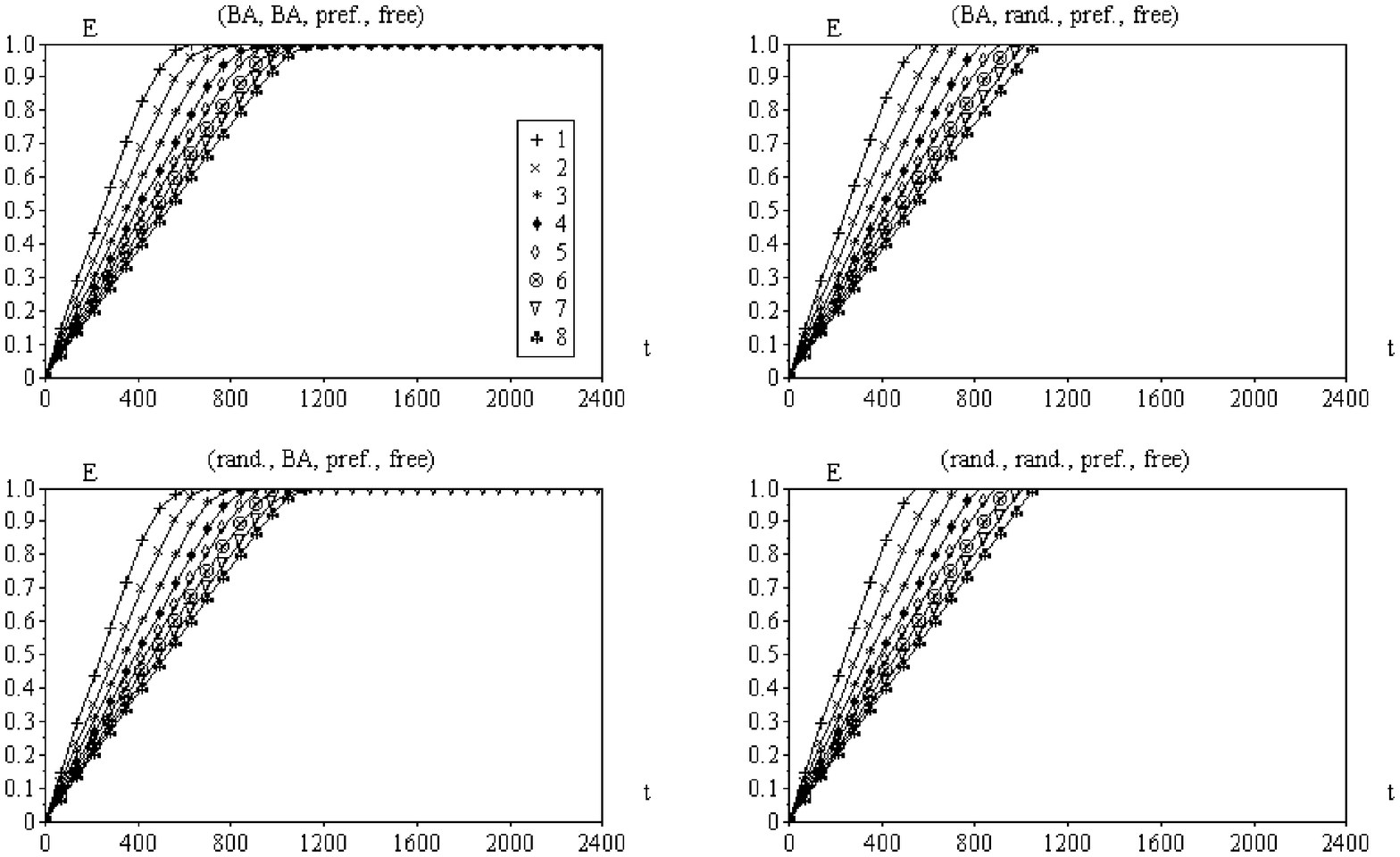} \\
   \vspace{0.5cm} 
   \caption{The percentage of visited edges $E$ in terms of the time
   $t$ for the configuration involving preferential choice of edges
   and unconditional interconnections.~\label{fig:E_pref_uncond}}
\end{center}
\end{figure*}

The percentages of visited edges $E(t)$ at each time instant are given
in Figures~\ref{fig:E_rand_cond} (random movements, conditional
interconnections), ~\ref{fig:E_rand_uncond} (random movements,
unconditional interconnections), ~\ref{fig:E_pref_cond} (preferential
movements, conditional interconnections) and ~\ref{fig:E_pref_uncond}
(preferential movements, unconditional interconnections).

\section{Discussion}

The results presented in the previous section are discussed in the
following with respect to the two main performance situations
considered in this work: number of visited nodes and edges.

\subsection{Knowledge in terms of Visited Nodes}

\emph{Effects of conditional interconnections} 
Compared to unconditional interconnections, the case of conditional
interconnections tends to substantially reduce the knowledge
acquisition speed.  This was expected indeed, because the conditional
interconnections imply the moving agent to stay longer wandering at
previous layer networks in order to collect the tokens necessary to
proceed into new layers.

\emph{Effects of the network models} As can be easily 
inferred by comparing the left and right columns of
figures~\ref{fig:rand_cond}-~\ref{fig:pref_uncond}, interconnections
through BA subnetworks have about the same effect as random networks
on the knowledge acquisition in all cases.  This is mainly a
consequence of the imposed similar connectivity of the BA and random
counterpart models used for the interconnecting subnetworks.
Similarly, the use of BA or random models for the layer networks also
led to minimal effect on the knowledge acquisition dynamics.  In
brief, the type of network model, BA or random, had little effect on
the overall knowledge acquisition efficiency.

\emph{Effects of the density of interconnections} Denser
interconnecting subnetworks tend to decrease the knowledge acquisition
speed in the case of conditional interconnections, having little
effect for unconditional interconnections (i.e. the learning curves
are nearly identical whatever the interconnecting density in
Figures~\ref{fig:rand_uncond} and ~\ref{fig:pref_uncond}).  Such a
behavior is explained because a larger number of conditional
interconnections implies the moving agent to collect more tokens in
the previous layers before proceeding to further layers.

\emph{Presence of Plateaux} The preferential movement strategy defined
for conditional interconnections has implied a series of plateaux of
knowledge acquisition along the learning curves.  The learning curves
in Figure~\ref{fig:pref_cond} are characterized by being preceeded by
a quick acquisition stage, followed by the respective plateau, whose
width tends to become larger as time goes by.  These plateaux indicate
a phase of knowledge stagnation, corresponding to the state of
dynamics of the system where the walks proceed predominantly over
edges in the previous layers, while the conditional links leading to
the subsequent layers are not yet feasible.  With this respect, it is
possible to draw a na\"{\i}ve analogy with a particle moving along a
series of chambers limited by successive compartments which are
progressively removed.  Congruently, the plateaux tend to become
larger along time because the walks have each time more alternatives
of random movement among the feasible edges.  This possibility is
corroborated by the fact that the plateaux become more discernible for
large interconnectivities (i.e. large values of $m$ adopted for the
interconnections in BA and random counterparts), which imply more
edges between subsequent layers.  The (possibly counterintuitive)
tendency of the preferential movements to reduce the knowledge
acquisition rate when compared to the random strategy can be explained
by the fact that in the preferential case the agent is forced to waste
time going through untracked edges in both layer and interconnecting
networks even cases where most adjacent nodes have already been
visited.  

\begin{figure*}
 \begin{center} 
   \includegraphics[scale=0.62,angle=0]{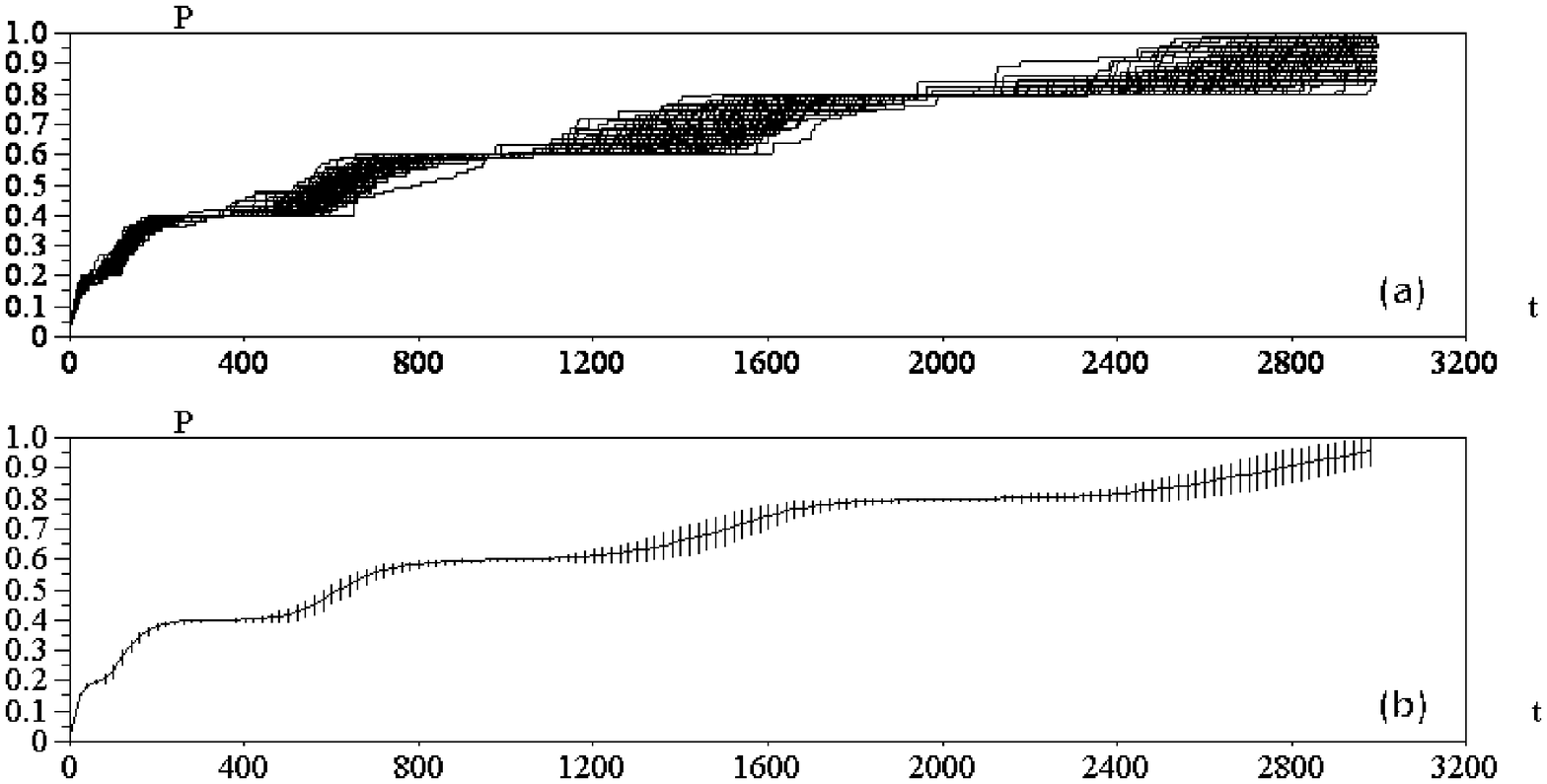} \\
   \vspace{0.5cm} 
   \caption{Visualization of 100 different trajectories obtained for
   5 layers with $N=20$ and $m=10$, interconnected through networks
   with $m=8$ (a), and respective average and standard deviations (b).
   ~\label{fig:traj_devs}}
\end{center}
\end{figure*}

An important issue related to the characterization of such plateaux is
whether the several individual trajectories obtained during simulation
are well represented in terms of their respective average values.
Figure~\ref{fig:traj_devs}(a) shows 100 such different trajectories
obtained for 5 layers with $N=20$ and $m=10$, interconnected through
networks with $m=8$. The average and standard deviations of these
trajectories are shown in Figure~\ref{fig:traj_devs}.  It is clear
from these figures that the trajectories do present small dispersion,
being therefore properly represented in terms of the respective
average.  Note also the first short plateau with height 0.2, which was
not visible in the previous figures because of the smaller resolution
of those pictures.

\begin{figure*}
 \begin{center} 
   \includegraphics[scale=0.62,angle=0]{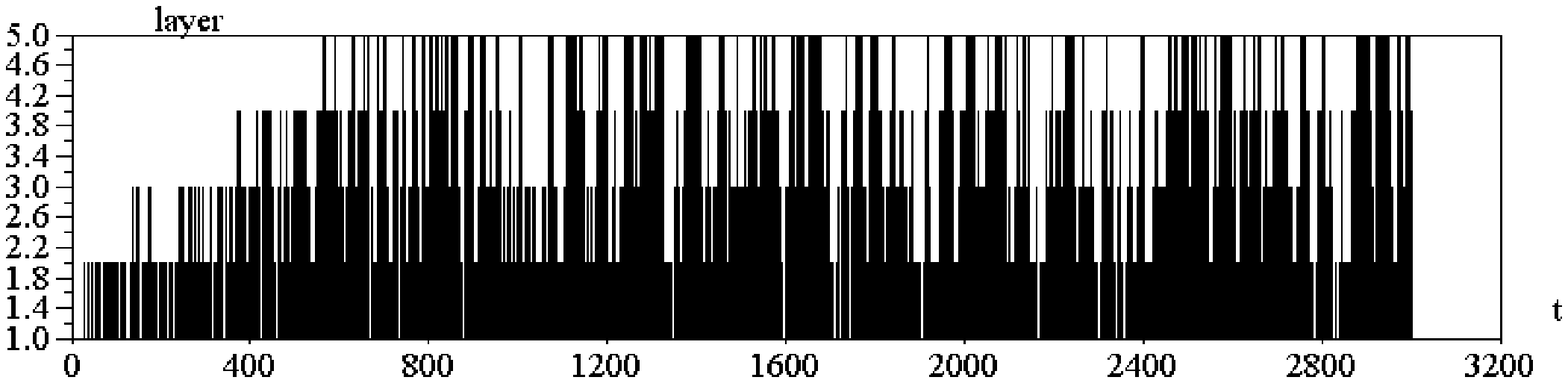} \\
   (a) \\
   \vspace{0.4cm}
   \includegraphics[scale=0.62,angle=0]{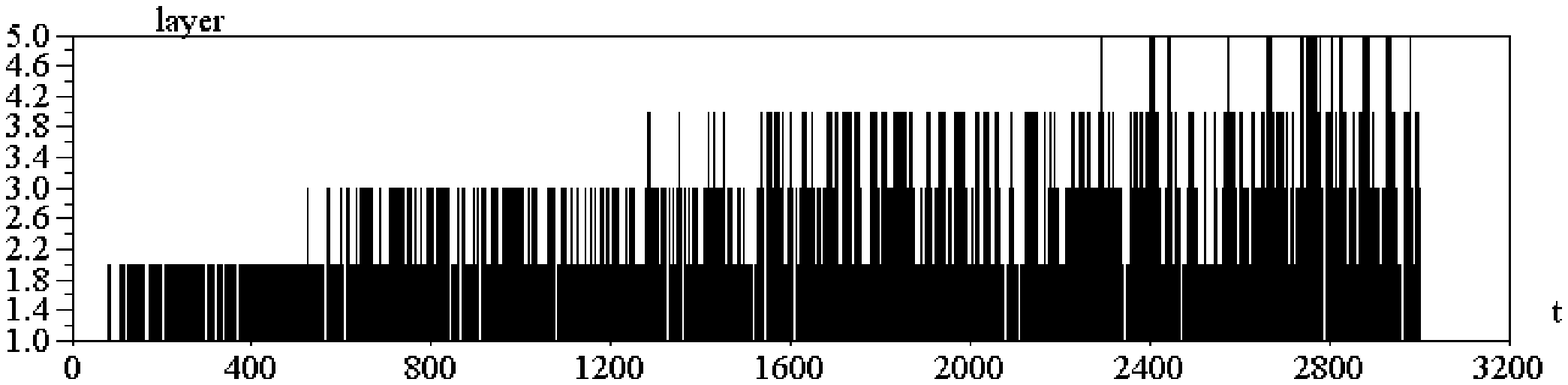} \\ 
   (b) \\
   \vspace{0.5cm} 
   \caption{The number of layer occupied by the moving agent in terms
   of the random walk steps $t$ for specific realizations considering
   random (a) and preferential (b) movements.~\label{fig:layers}}
\end{center}
\end{figure*}

Additional insight about the evolution of the ratio of visited nodes
in the presence of conditional edges can be obtained by considering
the number of the layer visited by the agent along the random walk
steps.  Such a curve is illustrated in Figure~\ref{fig:layers} for
specific realizations considering random (a) and preferential (b)
agent movements.  It is clear from this figure that the preferential
random walk implies the agent to explore most of the nodes in the
current layer, while seeking for free edges, before proceeding to
explore the subsequent layers, therefore implying the formation of
plateaux.

\emph{Layer Networks with other Connectivities}  In order to investigate 
the effect of the connectivity of the layer networks on the overall
knowledge acquisition dynamics, the above simulations were performed
also for $m=1$ and $10$ (all other situations discussed in this
subsection refer to layer networks with $m=5$).  It is clear from
Figures~\ref{fig:pref_cond_1} and~\ref{fig:pref_cond_10} that the
larger number of edges in each layer implied by $m=10$ tends to
substantially slow down the node coverage and to yield more marked
plateaux.

The overall fastest knowledge acquisition was observed for the cases
involving free transitions, with some speed up verified for the
preferential movement strategy (i.e. Figure~\ref{fig:pref_uncond}).

\subsection{Knowledge in terms of Visited Edges}

The dynamics of knowledge acquisition can also be quantified in terms
of the percentage $E(t)$ of \emph{visited edges}, which provides
additional insights about the considered models and strategies.  These
results are shown in Figures~\ref{fig:E_rand_cond}
to~\ref{fig:pref_uncond}.  The curves obtained for random movements
(i.e. Figures~\ref{fig:E_rand_cond} and~\ref{fig:E_rand_uncond})
are quite similar, indicating that the presence of conditional
edges has little effect 0n the edge coverage under the random movement
strategy.  The results obtained for preferential movements and
conditional transitions (i.e. Figure~\ref{fig:pref_cond}) indicate
that the edges are covered less effectivey than in the two previous
cases, especially for denser interconnections.  The fastest coverage
of edges was clearly obtained for preferential movements with free
transitions (i.e.~\ref{fig:E_pref_uncond}), which is a direct
consequence of the preference for new edges imposed by that strategy.
Such a fast edge coverage is also accompanied by the fastest node
coverage in Figure~\ref{fig:pref_uncond}.  Also interesting is the
fact that, though the edge coverage obtained for random movements
(i.e. Figures~\ref{fig:E_rand_cond} and~\ref{fig:E_rand_uncond})
resulted quite similar, the node coverage was verified to be much
faster in the former situation than in the latter.  Actually, the case
involving random movements and conditional transitions is
characterized by the fact that most nodes are covered after
approximately 1500 basic time steps (see Figure~\ref{fig:rand_cond})
even though only a fraction of respective edges have been covered at
that time, as indicated by Figure~\ref{fig:E_rand_cond}.

In order to better understand the preferential dynamics, let us first
consider its initial stages, where the agent starts its exploration of
the first and second layers.  Because preference is given to free
untracked edges, the agent tends to remain in layer 1 until most of
its free edges are tracked.  At this point, not only few free edges
remain untracked, but also most conditional links have been enabled.
Therefore, in presence of few untracked free edges, the agent
considers more frequently movements going through tracked or enabled
conditional edges, the latter leading to higher hierarchies. At
subsequent stages, when the agent is exploring a higher hierarchy, it
will tend to go through the free untracked edges, which now include
the conditional edges leading back to previous layers, proceeding into
higher layers only when the edges within and between the previous
layers have been mostly covered, so that few preferential movements
are allowed and the agent now considers more frequently going through
already tracked links on the current or previous layers or enabled
conditional links leading to higher layers.

The lack of plateaux in traditional random walks in presence of
conditional links can now more easily understood as being a
consequence of the fact that, by treating free or enabled conditional
edges with the same priority, irrespectively of being already traked
or not, allows more frequent explorations of the enabled conditional
edges leading to higher hierarchies.  Note that even at the earliest
stages of the exploration depicted in Figure~\ref{layers}(a), the
agent manages to get as far as the last layer.  In this way, plateaux
of stagnation are completely avoided.  However, it remains an
interesting fact that the barriers of conditional links are overcomed
with relative ease by the moving agent.

\section{Analytical Model} \label{sec:analytic}

The several interesting dynamical features so far identified through
numerical simulations are investigated further, especially regarding
their behavior under scaling of the network sizes and number of
layers, in the present section through a simplified analytical model.
Though we limit this investigation to preferential random walks in the
networks, more specifically the case leading to plateaux, the other
situations considered in this article can be treated similarly.

We start by considering the fact that, at step $t$ of a random walk
preferential to untracked edges, the ratio of visited nodes $P(t)$ of
a random or BA network with $N$ nodes and free links has been been
verified, through extensive simulations, to be approximated (at least
for $10 \leq N \ 1000$) as

\begin{equation}
  P(t) = 1 - e^{-t/N} \label{eq:p}
\end{equation}

We shall make a small modification of the way in which the conditional
edges are considered so as to simplify the analytical characterization
of the knowledge acquisition dynamics.  More specifically, starting at
layer $h=1$, movements to the subsequent layer $h+1=2$ will only be
allowed after a ratio $P_c$ of visited nodes in $h$ has been achieved.
The difference between such an assumption and the situations so far
considered in the present article is that in the latter situation the
moving agent is allowed to explore subsequent layers at any time,
provided it holds the respective prerequisites.  However, the above
simplification holds particularly well when the connectivity between
subsequent layers is relatively large with respect to the connectivity
between the nodes in each layer, because in such a situation most
movements in the subsequent layer will be mostly blocked by the
prerequisites and preference to free untracked edges in the current
and previous layers.

Let $\left< k \right>$ be the average degree at each layer and $\left<
k_i \right>$ be the average degree of the interconnecting layer.  At
the beginning of the random walk, the exploration is limited to layer
$h=1$, so that $P_c$ is reached at a critical step $t_c$ so that
$P(t_c)=P_c$ which can be calculated through Equation~\ref{eq:p}.

\begin{figure*}
 \begin{center} 
   \includegraphics[scale=0.62,angle=0]{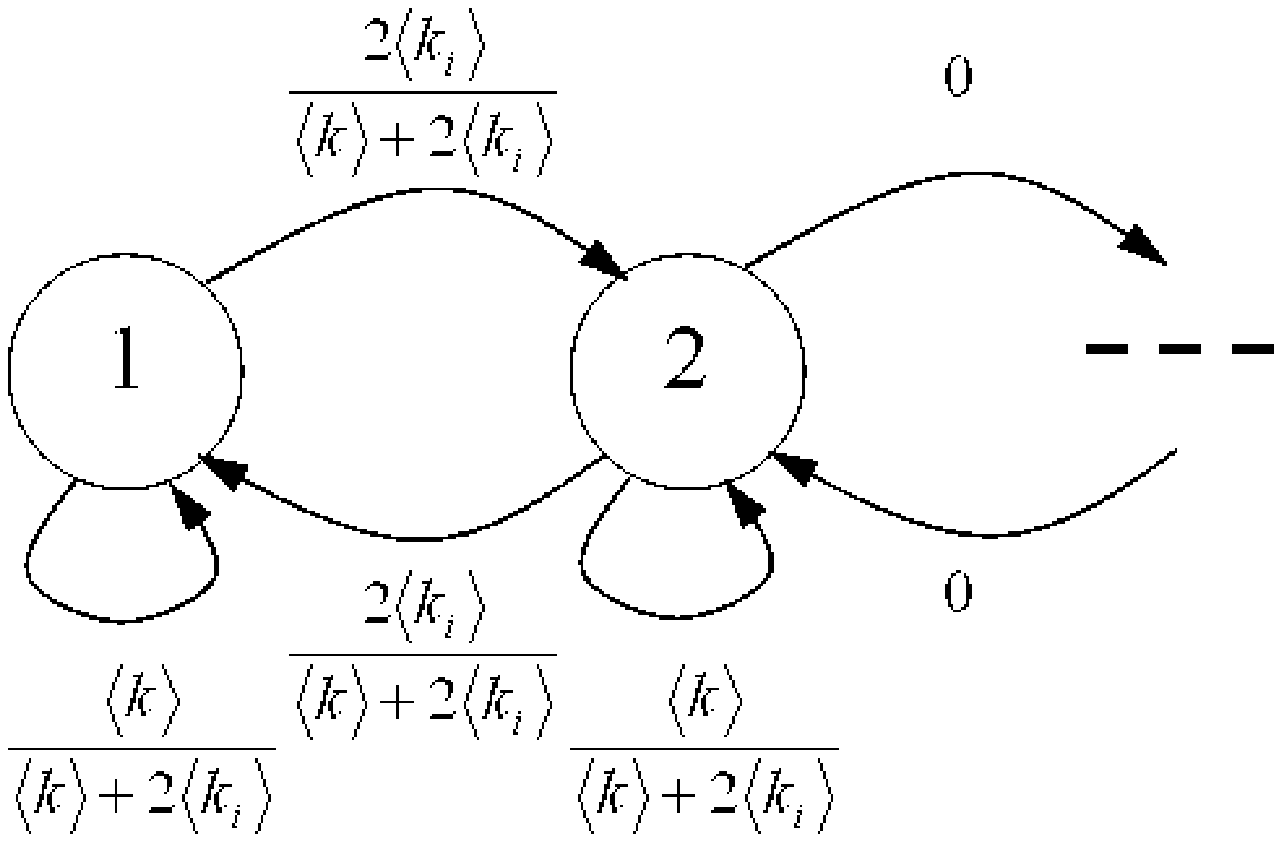} \\
   (a) \\
   \vspace{0.4cm}
   \includegraphics[scale=0.62,angle=0]{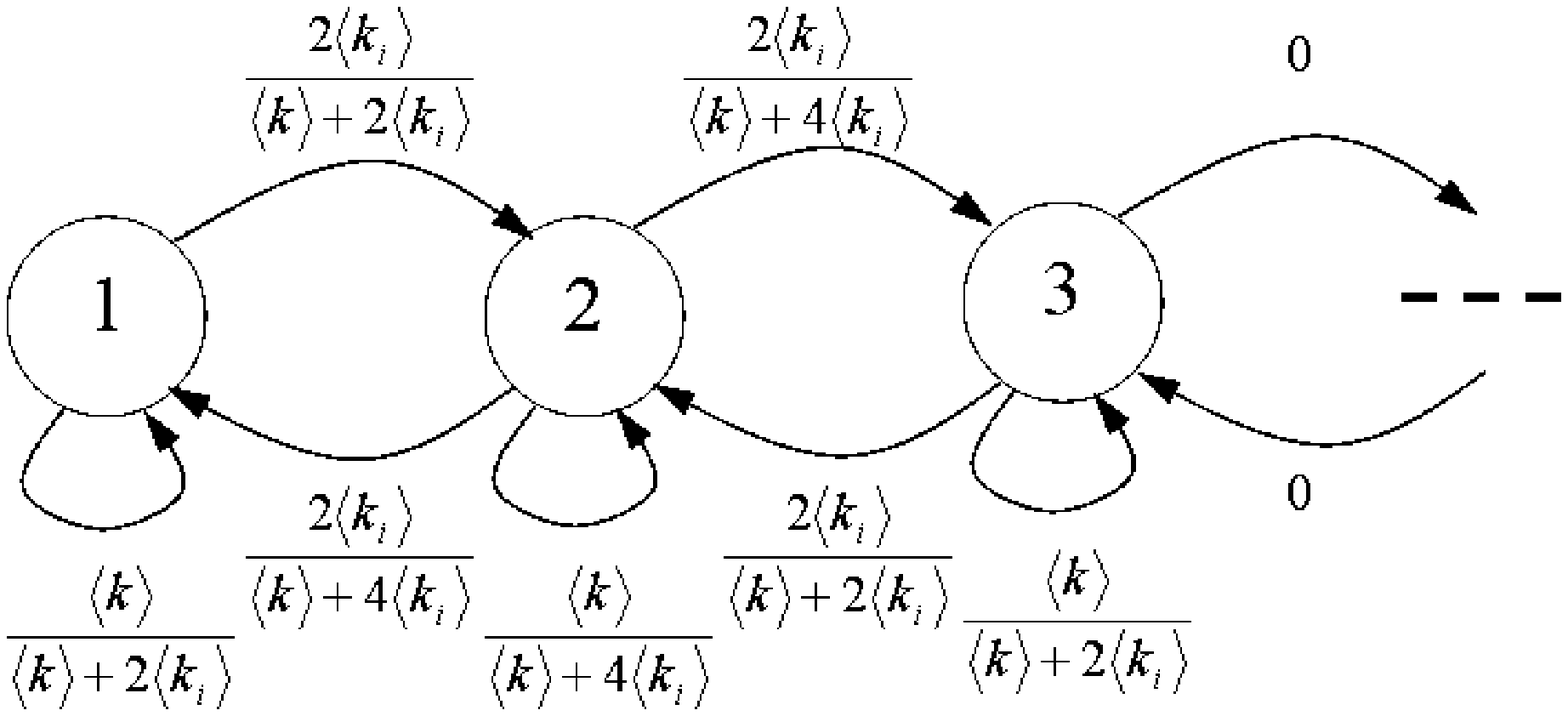} \\
   (b) \\
   \vspace{0.5cm} 
   \caption{The movement of the agent between subsequent
   hierarchical layers can be modeled in terms of Markovian
   models, as illustrated for two (a) and three (b) layers.
   Note that the progress to the subsequent layers are blocked
   until the critical ratio of visited nodes $P_c$ is achieved
   for the last enabled layer.~\label{fig:markov}}
\end{center}
\end{figure*}

Afterwards, all conditional links at layer $2$ are enabled, so that
the exploration of that layer begins.  However, as the interconnecting
edges are bidirectional, the agent will now exchange between layers
$1$ and $2$ until the ratio $Pc$ is achieved for layer $2$.  The
respective occupancy of these two layers can be approximated in terms
of the Markov chain shown in Figure~\ref{fig:markov}.  

The stochastic matrix $S$ associated to this Markovian system is
immediately obtained as

\begin{equation}
  S_2=\left[ \begin{array}{cc}
     \frac{\left< k \right>}{\left< k \right>+2\left< k_i \right>}  &
       \frac{2\left< k_i \right>}{\left< k \right>+2\left< k_i \right>} \\
     \frac{2\left< k_i \right>}{\left< k \right>+2\left< k_i \right>} &
       \frac{\left< k \right>}{\left< k \right>+2\left< k_i \right>}  
  \end{array}
  \right]
\end{equation}

where the factor 2 stands for the fact that one undirected edge in the
interconnecting layer implies two conditional links between layer $1$
and $2$ (see Figure~\ref{fig:interconn}).

Although the moving agent will soon be spending the same proportion of
time at layers 1 or 2, it is the relative frequency $F$ of time steps
at which the moving agent remains at or enters into layer $2$ that
matters for the coverage of the nodes in this layer and respective
acquisition of prerequisites.  This frequency is immediately given as
being equal to the relative number of movements through the two edges
leading to node $2$.  Tt follows by symmetry that $F=0.5$, implying
the exploration of layer $2$ to be effectively performed at a
renormalized $P(t)$ given as

\begin{equation}
  P(t) = 1 - e^{- 0.5 t /N}  \label{eq:p2}
\end{equation}

Note that $t$ is relative to each exploration stage, always starting
when the critical ratio of visited nodes is achieved for the last
current layer.  After liberating layer $3$ for exploration, the Markov
model becomes as shown in Figure~\ref{fig:markov}(b), and the
respective stochastic matrix now reads

\begin{equation}
  S_3=\left[ \begin{array}{ccc}
     \frac{\left< k \right>}{\left< k \right>+2\left< k_i \right>}  &
       \frac{2\left< k_i \right>}{\left< k \right>+4\left< k_i \right>} &
       0  \\
     \frac{2\left< k_i \right>}{\left< k \right>+2\left< k_i \right>} &
       \frac{\left< k \right>}{\left< k \right>+4\left< k_i \right>} &
       \frac{2\left< k_i \right>}{\left< k \right>+2\left< k_i \right>} \\ 
     0 &
       \frac{2\left< k_i \right>}{\left< k \right>+4\left< k_i \right>}  &
       \frac{\left< k \right>}{\left< k \right>+2\left< k_i \right>} \\
  \end{array}
  \right]
\end{equation}

Note that all subsequent stochastic matrices will share the right-hand
lower $3 \times 2$ block with the above matrix, from which a generic
probabilistic model can be developed.  At a generic hierarchical level
$h$, these four elements are given as follows

\begin{eqnarray}
  S_{h-2,h-1}=\frac{2\left< k_i \right>}{\left< k \right>+4\left< k_i \right>}  \label{eq:Sh2h1}\\
  S_{h-1,h-1}=\frac{\left< k \right>}{\left< k \right>+4\left< k_i \right>} \\
  S_{h-1,h}=\frac{2\left< k_i \right>}{\left< k \right>+2\left< k_i \right>} \\
  S_{h,h-1}= S_{h-2,h-1} \\
and \nonumber \\  
  S_{h,h}=\frac{\left< k \right>}{\left< k \right>+2\left< k_i \right>} \label{eq:Shh}
\end{eqnarray}

Because of the inherent symmetry of the transition probabilities in
the matrix $S$, the occupancy $p(h)$ of each state $h$ can be
calculated as

\begin{equation}
     p(h) = \left\{ \begin{array}{ll}
               \frac{1}{2+(h-2)(1+S(h-1,h))}  &  \mbox{for $h=1$ or $h=H$} \\
                \\
               \frac{1+S(h-1,h)}{2+(h-2)(1+S(h-1,h))}  &  \mbox{otherwise}
                    \end{array}
            \right.   \label{eq:ph}
\end{equation}

As the relative frequency in which each of these transitions $S(a,b)$
are performed is immediately given as $p(b)S(a,b)$, we have that the
relative frequency of movements $F(h)$ into the last currently enabled
layer $h$ is therefore given as

\begin{eqnarray}
  F(h) = \frac{\alpha}{\beta+(h-2)\gamma}    \label{eq:F}   \\
  \mbox{where}  \nonumber \\
  \alpha = p(h-1)S_{h,h-1}+p(h)S_{h,h} \nonumber  \\
  \beta = 2(p(h)S_{h-1,h}+p(h)S_{h,h}) \nonumber \\
  \gamma = 2p(h-1)S_{h-2,h-1}+p(h-1)S_{h-1,h-1}  \nonumber
\end{eqnarray}

By substituting Equations~\ref{eq:Sh2h1}--\ref{eq:Shh} and~\ref{eq:ph}
into Equation~\ref{eq:F}, it follows that

\begin{eqnarray}
  F(h) = \frac{A}{B+hC}    \label{eq:Fh}   \\
  \mbox{where}  \nonumber \\
  A = \left< k \right>^2 + 6 \left< k \right> \left< k_i \right> + 8 
           \left< k_i \right>^2  \nonumber \\
  \mbox{and} \nonumber \\
  B = -4 \left< k_i \right>\left< k \right> -16 \left< k_i \right>^2 
                 \nonumber \\
  C = \left< k \right>^2+8 \left< k_i \right>\left< k \right> +
           16 \left< k_i \right>^2  \nonumber 
\end{eqnarray}

The evolution of the ratio of visited nodes can therefore be estimated
by using Equations~\ref{eq:p} (for $h=1$) and~\ref{eq:pgen} (for $h
\geq 2$).

\begin{equation}
  P(t) = 1 - e^{- F(h) t /N}  \label{eq:pgen}
\end{equation}

This implies the overall evolution to be composed by subsequent time
scaled versions of Equation~\ref{eq:p}, given by
Equation~\ref{eq:pgen} in terms of the value $F(h)$.  Consequently,
the length of each stage along the liberation of the layers will be
given as

\begin{equation}
  L(h)= 1/F(h) = \frac{B+hC}{A}
\end{equation}

where $A$, $B$ and $C$ are constants defined by $\left< k \right>$ and
$\left< k_i \right>$.  It is now clear that this length scales
proportionally to $h$.  Equation~\ref{eq:pgen} also provides the means
for analyzing the scaling of $P(t)$ with $N$.  Because the coefficient
$c$ of the exponential in $P(t)$, i.e. $c=F(h)/N$, corresponds to a
product between $F(h)$ and $1/N$, the scaling of the subnetworks size
from $N$ to $aN$ will imply $\tilde{c} = (F(h)/a) N = c/a$ for all
$h$, i.e. the length of all stages $\tilde{L}(h)$ will be equal to
$aL(h)$.  In other words, the overall shape of the ratio of visited
nodes will not change when $N$ is scaled while all other parameters
are kept fixed.

\begin{figure*}
 \begin{center} 
   \includegraphics[scale=0.62,angle=0]{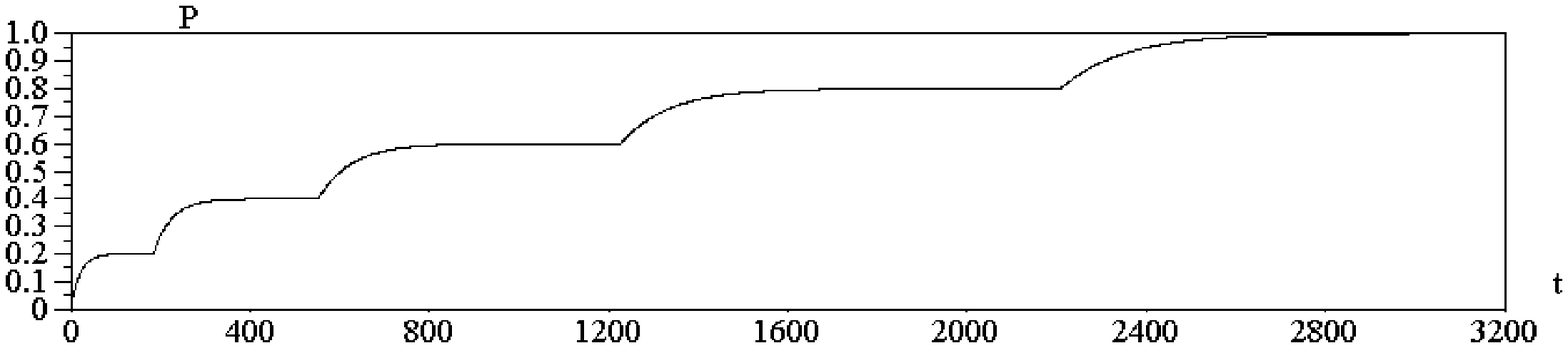} \\
   \vspace{0.5cm} 
   \caption{Analytical evolution of the ratio of visited nodes
   for a random network with $N=20$, $\left< k \right>=10$, $\left< k_i
   \right>=16$, $P_c=0.9999$, and five layers.~\label{fig:analytic}}
\end{center}
\end{figure*}

Figure~\ref{fig:analytic} illustrates the evolution of the ratio of
visited nodes in a BA network as estimated by the above model assuming
$N=20$, $\left< k \right>=10$, $\left< k_i
\right>=16$, $P_c=0.9999$, and five layers.  Except for the value
$P_c$, these parameters correspond precisely to those considered in
the evolution shown in Figure~\ref{fig:traj_devs}.  The specific value
of $P_c$ was chosen so as to obtain proper fitting between the
experimental data and the theoretical model.  A good overall adherence
can be observed between the analytical and respective experimental
evolutions regarding both the lengths and heights of each plateau.
Interestingly, the analytical model also captures the fact that
progressively smoother transitions are obtained at higher
hierarchies. The main difference between these evolutions are related
to the the fact that, in the original experiment, the moving agent was
allowed to proceed to subsequent layers more freely, i.e. before the
critical ratio of visited nodes $P_c$ had been reached.  Such a
dynamics would contribute to smoothing the lenf-hand side of the
evolution curve at each transition, as is the case in the experimental
results in Figure~\ref{fig:traj_devs}.

\section{Concluding Remarks}

This article has presented a simple approach to knowledge acquisition
based on representation of knowledge as a hierarchical complex
network~\cite{Costa_vor:2003} and the modeling of the process of
knowledge acquisition in terms of walks along such networks.  Though
simple, the considered models incorporate the existence of two types
of edges (free and conditional), including multiple conditional
transitions where the access to specific nodes are granted only after
the agent has visited specific nodes.  This movement strategy
represents a possibly new mechanism for complex network and random
walk researches.

Two visiting strategies have been considered: at random and
preferential to still untracked free edges.  Simulations considering
several densities of connectivity between 5 hierarchical layers have
been evaluated with respect to conditional interconnecting networks
and unconditional counterparts, and the knowledge acquisition dynamics
quantified in terms of the number of visited nodes and edges as a
function of time (i.e. each basic movement of the agent).  A Markovian
analytical model of the learning dynamics is developed for the case of
preferential random walks in presence of conditional links, which
reproduces the plateaux heights and lengths.  This model has allowed
the discussion of the scaling properties of the dynamics with respect
to the network size and number of layers.  Among other findings, the
lengths of the plateaux have been verified to be proportional to the
number of already explored layers.

Despite the simplicity of the approach, a series of interesting
complex dynamics and effects have been identified from the learning
curves, including the fact that the preferential movement strategy
was slower than the random counterpart for the case of
conditional interconnections, as well as the identification of
plateaux of stagnation of learning for the latter situation.

The reported work has paved the way to several future works, including
the consideration of multiple agents~\cite{Acedo_Yuste:2003}, which
may or may not share information about their individual adjacency
matrices.  Another relevant issue to be incorporated into the model is
the fact that the transitions from one node to another, i.e. the
inference of some subset of knowledge from another, may not always
take the same time.  It would therefore be interesting to consider
diverse distributions of time-weights along the hierarchical knowledge
networks. Also interesting is the fact that the suggested approach and
models provide an interesting framework for investigating data flow
architectures (e.g.~\cite{Silc:1999}).  This type of parallel
computing architecture is characterized by a hierarchical processing
flow constrained by dependences between intermediate computing stages,
which could be conveniently modeled by the hierarchical complex
networks with multiple conditional edges. It would be interesting to
consider additional measurements typically used in random walk
investigations, such as return time and correlations.

\vspace{1cm}

Luciano da F. Costa is grateful to Dietrich Stauffer, Osvaldo Novais
de Oliveira Jr. and Gonzalo Travieso for careful reading and
commenting on this article, and to FAPESP (process 99/12765-2) and
CNPq (308231/03-1) for financial sponsorship.

\bibliographystyle{apsrev}
\bibliography{know}
\end{document}